\documentclass[twocolumn]{aastex63}
\usepackage{amsmath,fancyvrb}	
\usepackage{amssymb}	
\usepackage{array}
\usepackage{bm}
\usepackage[utf8]{inputenc}
\usepackage{CJK}
\usepackage{graphicx}	
\usepackage{xifthen}
\usepackage{xspace}
\usepackage{multirow}

\newcommand{\tablefoot}[1]{\par\smallskip\footnotesize #1}

\DeclareRobustCommand{\zh}[1]{%
  \begin{CJK*}{UTF8}{gbsn}\ignorespaces#1\unskip\end{CJK*}%
  \kern-0.30em%
  }

\begin{document}

\title{Multi-wavelength ALMA Imaging of HD 34282: Dust-trapping Signatures of a Vortex Candidate}

\author[0000-0003-2622-6895]{Xiaoyi Ma (\zh{马潇依})}
\affiliation{Kavli Institute for Astronomy and Astrophysics, Peking University, Beijing 100871, China}
\affiliation{Department of Astronomy, School of Physics, Peking University, Beijing 100871, China}

\author[0009-0004-6973-3955]{Fangyuan Yu (\zh{俞方远})}
\affiliation{Department of Astrophysical Sciences, Princeton University, 4 Ivy Lane, Princeton, NJ 08544, USA}

\author[0000-0001-9290-7846]{Ruobing Dong (\zh{董若冰})}
\affiliation{Kavli Institute for Astronomy and Astrophysics, Peking University, Beijing 100871, China}

\author[0000-0003-1958-6673]{Kiyoaki Doi}
\affiliation{Max-Planck-Institut fur Astronomie, Konigstuhl 17, 69117 Heidelberg, Germany}

\author[0000-0003-4562-4119]{Akimasa Kataoka}
\affiliation{National Astronomical Observatory of Japan, 2-21-1 Osawa, Mitaka, Tokyo 181-8588, Japan}

\author[0000-0003-2300-2626]{Hauyu Baobab Liu}
\affiliation{Department of Physics, National Sun Yat-Sen University, No. 70, Lien-Hai Road, Kaohsiung City 80424, Taiwan, R.O.C.}
\affiliation{Center of Astronomy and Gravitation, National Taiwan Normal University, Taipei 116, Taiwan}

\author[0000-0002-7607-719X]{Feng Long (\zh{龙凤})}
\affiliation{Kavli Institute for Astronomy and Astrophysics, Peking University, Beijing 100871, China}

\author[0000-0003-4902-222X]{Takahiro Ueda}
\affiliation{National Astronomical Observatory of Japan, 2-21-1 Osawa, Mitaka, Tokyo 181-8588, Japan}

\author[0009-0003-0553-7456]{Huojun Li (\zh{李火骏})}
\affiliation{Department of Astronomy, School of Physics, Peking University, Beijing 100871, China}

\author[0000-0003-2458-9756]{Nienke van der Marel}
\affiliation{Leiden Observatory, Leiden University, P.O. Box 9513, 2300 RA Leiden, The Netherlands}

\author[0000-0001-7157-6275]{\'Agnes K\'osp\'al}
\affiliation{Konkoly Observatory, HUN-REN Research Centre for Astronomy and Earth Sciences, MTA Centre of Excellence, Konkoly-Thege Mikl\'os \'ut 15-17, 1121 Budapest, Hungary}
\affiliation{Max-Planck-Insitut f\"ur Astronomie, K\"onigstuhl 17, 69117 Heidelberg, Germany}

\correspondingauthor{Xiaoyi Ma, Ruobing Dong}
\email{xiaoyima@stu.pku.edu.cn, rbdong@pku.edu.cn}

\begin{abstract}

Azimuthal arcs in millimeter continuum emission from protoplanetary disks are often attributed to dust-trapping vortices, but definitive observational confirmation of vortices remains lacking. We present sub-$0.1''$ resolution ALMA continuum observations of the HD~34282 disk at 0.9, 1.3, 2.1, and 3.1~mm. These observations resolve a bright azimuthal arc superposed on a compact double-gap, triple-ring morphology, most clearly at shorter wavelengths, and enable us to probe the physical origin of the arc. It exhibits a lower spectral index than the surrounding rings, consistent with enhanced grain growth and/or higher dust surface density of a dust-trapping vortex. Its azimuthal width decreases with increasing wavelength, consistent with tighter confinement of larger grains, or lower optical depths at longer wavelengths. These observations probe dust with Stokes numbers $\mathrm{St}\lesssim0.03$. Vortex models predict negligible peak shifts in this regime, consistent with the 1.3–3.1~mm data. At 0.9~mm, however, the arc peak is offset by $\sim15^\circ\pm4^\circ$ in the direction of disk rotation relative to longer wavelengths, and the near-side ring emission is locally dimmer compared to the far-side, likely reflecting optical-depth or temperature effects. These observations are consistent with azimuthal dust trapping, potentially associated with a vortex-induced pressure maximum.
\end{abstract}

\section{Introduction}
High-resolution observations of protoplanetary disks reveal that substructures are ubiquitous \citep[e.g.,][]{Andrews_2018, Long_2018}. Many systems exhibit axisymmetric rings and gaps in disks, while a subset shows striking azimuthal arcs in the millimeter continuum \citep[e.g.,][]{van_der_Marel_2021, Curone_2025}. Millimeter continuum arcs are typically observed at radii of tens to hundreds of au and span a finite azimuthal extent, ranging from $20^\circ$ to $170^\circ$, often well shorter than a full ring \citep[e.g.,][]{van_der_Marel_2013, Casassus_2015, van_der_Marel_2016, Dong_2018, van_der_Marel_2021, Curone_2025}. When resolved, they are radially compact and are often embedded within ring structures, suggesting localized dust enhancements \citep{van_der_Marel_2015b}. A range of physical mechanisms have been proposed to explain azimuthal arcs, and their origins are not uniquely constrained.

A leading explanation for the observed arcs is azimuthal dust trapping, potentially associated with long-lived anticyclonic vortices, arising from the Rossby-wave instability (RWI) at sharp vortensity gradients in low-turbulence disks \citep{Lovelace_1978, Lovelace_1999, Li_2000, Ono_2018}, 
such as at planet-carved gap edges \citep{Zhu_2014a, Zhu_2014b, Cimerman_2023} or viscosity transitions \citep{Regaly_2012, Flock_2015}. These vortices act as local high-pressure maxima that trap dust both azimuthally and radially \citep{Lyra_2008, Lin_2012, Baruteau_2016}. By concentrating dust, they increase local dust-to-gas ratio, which may enable the streaming instability \citep{Stammler_2019}. This accelerates grain growth and coagulation within the vortex \citep{Birnstiel_2013, Fu_2014}, making vortices plausible planetesimal birth sites \citep{Meheut_2012}. Moreover, vortices can also sculpt rings and gaps by launching density waves \citep{Paardekooper_2010}, further trapping dust, which may facilitate later stages of planet formation \citep{Ma_2025}. Identifying these features in disks is therefore crucial for linking observed substructures to the mechanisms that govern planet formation.

In addition to continuum observations, vortices may also be revealed by scattered-light imaging that traces micrometer-sized grains on disk surfaces. Simulations predict that vortices appear as bright single-arm spirals and cast shadows \citep{Marr_2022} in scattered-light images for moderately inclined disks. Among disks with mm-continuum arcs, only a few have high-sensitivity, high-resolution scattered light data, such as HD~34282 \citep{van_der_Plas_2017, de_Boer_2021, Curone_2025}, HD 142527 \citep{Casassus_2015, Marino_2015a}, HD 135344B \citep{Stolker_2016, van_der_Marel_2016}, V1247 Ori \citep{Kraus_2017}, MWC 758 \citep{Boehler_2018, Dong_2018}, and AB Aurigae \citep{Tang_2017, Boccaletti_2020}. Only some of these systems show spiral-like scattered-light features that spatially coincide with the mm arc, together with an outer-disk shadow (e.g., HD~34282; \citealt{Marr_2022}).

However, definitive evidence of vortices in gas motion is still lacking, and alternative explanations to observed arcs have been proposed, such as pile-ups of material at apocenter in eccentric disks \citep{Ataiee_2013, Ragusa_2017, Ragusa_2020}. A vortex should exhibit distinct kinematic signatures in gas velocity fields compared to the eccentric disk scenario. To date, such signatures remain elusive. Non-Keplerian velocity perturbations consistent with pressure maxima have been reported in HD~142527 \citep{Boehler_2018, Yen_Gu_2020}, but they are not uniquely attributable to vortical motion \citep{Ragusa_2017, Boehler_2021}. More recent ALMA gas observations of disks with mm continuum arcs also did not reveal unambiguous vortex kinematic signatures \citep{Wolfer_2025,stadler_2026}.

In millimeter dust emission, multi-wavelength observations provide a powerful and accessible diagnostic of azimuthal dust trapping in vortices. Arcs are expected to narrow at longer wavelengths because larger grains are more efficiently trapped \citep{Birnstiel_2013}, as observed in IRS 48 \citep{van_der_Marel_2015b}, MWC 758 \citep{Marino_2015b}, HD 142527 \citep{Casassus_2015}, HD 135344B \citep{Cazzoletti_2018}, and PDS 70 \citep{Doi_2024}. However, this trend alone supports dust trapping and does not uniquely establish a vortex origin. Analytic steady-state vortex prescriptions \citep{Lyra_2013} link continuum morphology to vortex parameters and enable forward modeling under specific dust assumptions. For example, the aspect ratio and wavelength dependence of the arc azimuthal width in MWC 758 have been successfully modeled using these vortex prescriptions \citep{Casassus_2019}. Hydrodynamic simulations also predict a wavelength-dependent azimuthal shift of the dust trapping location in vortices \citep{Baruteau_2016}. In HD~135344B, multi-wavelength ALMA data show a shift, but in a direction inconsistent with theoretical expectations \citep{Cazzoletti_2018}.

We present new high-resolution, high-sensitivity ALMA continuum observations of the HD~34282 disk at 3.1, 2.1, and 0.9 mm, together with archival 1.3 mm data. These data allow us to test the wavelength-dependent emission properties within the azimuthal arc and assess whether it is consistent with dust trapping. HD 34282 is a Herbig Ae star located at a distance of 309$\pm$2 pc, with a stellar mass of approximately 1.5 $M_\odot$, age of $\sim$6.4 Myr, a luminosity of $\sim$13.6 $L_\odot$ \citep{Gaia_2023}. It hosts a transition disk with inclination $i \sim 60^\circ$ with a cavity of $\sim$80 au in millimeter dust observations \citep{van_der_Plas_2017}. The disk exhibits a scattered-light spiral co-spatial with the mm-bright arc and an outer-disk shadow, making the structure a strong vortex candidate \citep{de_Boer_2021, Marr_2022}. Existing $^{12}$CO and $^{13}$CO molecular line observations did not show distinctive kinematic signature of the vortex in HD 34282 likely due to insufficient sensitivity and resolution \citep{Law_2023, Wolfer_2025}.

The paper is organized as follows. \S\ref{sec:observation&reduction} describes the observations and data reduction processes. \S\ref{sec:result} presents the disk morphology and visibility modeling. \S\ref{sec:analysis} tests theoretical predictions of dust trapping against the observations. \S\ref{sec:discussion} discusses additional features at 0.9 mm beyond the rings and arc. Finally, \S\ref{sec:conclusions} summarizes our main findings.

\section{Observations $\&$ Data Reduction}\label{sec:observation&reduction}

In this section, we summarize the observations (\S\ref{sec:observation}) and data reduction (\S\ref{sec:data_reduction}). Table~\ref{table:observation} lists the key dataset and image characteristics.

\subsection{Observations}\label{sec:observation}
\begin{table*}
\begin{center} 
\caption{Observation and Image Summary}
\label{table:observation}
\begin{tabular}{ c c c c c c c } 
\hline
ALMA Band & Program ID  &  PI &  Obs.Freq & Synthesized Beam Size & RMS Noise & Peak SNR\\ 
 &   &  & [GHz] &  [arcsec] & [$\mu$Jy Beam$^{-1}$] & \\
 \hline \hline
Band 3 & 2022.1.00315.S & Ruobing Dong & 89.54 - 105.48  & 0.095$\times$ 0.081 & 5.99 & 60 \\
\hline
Band 4 & 2022.1.00315.S & Ruobing Dong & 137.04 - 152.99  & 0.081$\times$ 0.049 & 7.41 & 80 \\
\hline
\multirow{2}{*}{Band 6} & 2015.1.00192.S & Gerrit van der Plas & 215.94 - 235.18  & \multirow{2}{*}{0.056$\times$ 0.051} & \multirow{2}{*}{20.90} & \multirow{2}{*}{55 }\\
 & 2017.1.01578.S &  Jozua de Boer & 216.30 - 235.12  &  &  & \\
 \hline
\multirow{2}{*}{Band 7} & 2022.1.00315.S & Ruobing Dong & 335.55 - 351.50 & \multirow{2}{*}{0.059$\times$ 0.050} & \multirow{2}{*}{24.99} & \multirow{2}{*}{133}\\
 & 2023.1.00108.S & Xiaoyi Ma & 335.55 - 351.50  &  &  & \\
\hline
\end{tabular}
\tablefoot{%
\textbf{Note}: Column 1: ALMA Band; Column 2: ALMA program ID. For Band 3 and 4 observations, SB and LB executions come from the same ALMA program. For Band 6 and 7 observations, SB and LB come from different programs. Band 6 uses SB from 2015.1.00192.S and LB from 2017.1.01578.S. Band 7 uses SB from 2023.1.00108.S and LB from 2022.1.00315.S; Column 3: PI of the ALMA project; Column 4: Observation Frequency range covering all spectral windows; Column 5: Synthesized beam (FWHM) of the final image; imaging details in \ref{sec:data_reduction}; Column 6: RMS noise for the image; Column 7: Peak SNR.}
\end{center}
\end{table*}

We analyze HD~34282 using new ALMA continuum observations at 3.1, 2.1, and 0.9 mm (Bands 3, 4, and 7) together with archival 1.3 mm data (Band 6) \citep{Francis_2020} to characterize the disk asymmetry at high angular resolution and sensitivity. 

Band 3 observations were conducted in ALMA Cycle 10 between May and June 2023 as part of program 2022.1.00315.S (PI: R. Dong). The correlator was configured with four spectral windows in dual polarization and time division mode (TDM). Spectral windows are centered at 90.521, 92.416, 102.521, and 104.479 GHz with 128 channels each 15.625 MHz in width, spanning a total bandwidth of 2 GHz in each spectral window. The observations were carried out in both extended (long-baseline; LB) and compact (short-baseline; SB) configurations to achieve high spatial resolution while preserving large maximum recoverable scales for accurate total flux recovery. The SB observations used C-6 configuration of 42 antennas with baselines from 15.3 m to 2.5 km for a total on-source integration time of $\sim$49 min. The LB observations comprised four execution blocks with 47, 49, 47, and 48 antennas using C-9 configuration, covering baselines from 226.5 m to 15.2 km, with a total on-source integration time of $\sim$142 min. 

Band 4 observations were conducted in ALMA Cycle 10 between May and July 2023 as a part of program 2022.1.00315.S. The correlator was configured in dual polarization and TDM mode with the spectral windows centered at 138.021, 139.916, 150.021, and 151.979 GHz. The SB observations used C-6 configuration of 40 antennas with baselines from 15.2 m to 2.5 km for a total on-source integration time of $\sim$34 min. The LB observations comprised two execution blocks with 43 and 38 antennas using C-9 configuration, covering baselines from 113.3 m to 9.7 km, with a total on-source integration time of $\sim$82 min.

Band 6 data comprise archival observations, including SB executions from Cycle 3 (program 2015.1.00192.S, PI: G. van der Plas) and LB executions from Cycle 5 (program 2017.1.02578.S, PI: J. de Boer). We used the processed dataset in which the SB and LB executions were concatenated, as presented by \cite{Francis_2020}, which has SNR $\sim$55 and a synthesized beam of 0.056$\times$ 0.051$''$. The observational setup and calibration procedures are described in detail in \cite{Francis_2020}.

The Band 7 data included SB and LB observations executed from different programs. The LB observations at Band 7 were conducted on September 24, 2023 as part of program 2022.1.00315.S, with C-7 configuration of 41 and 45 antennas in two execution blocks covering baselines from 66.8 m to 5.8 km for a total on-source integration time of $\sim$74 min. The correlator was configured in TDM mode with spectral windows centered at 336.512, 338.416, 348.521, and 350.479 GHz. The SB observations were conducted in ALMA Cycle 11 on December 23, 2023 as a part of 2023.1.00108.S (PI: X. Ma) using 45 antennas with baselines from 15.1 m to 1.2 km using C-4 configuration for a total on-source integration time of $\sim$20 min. The correlator has the same spectral setup as LB observations. 

\begin{figure*}
\begin{center}
\includegraphics[width=\textwidth]{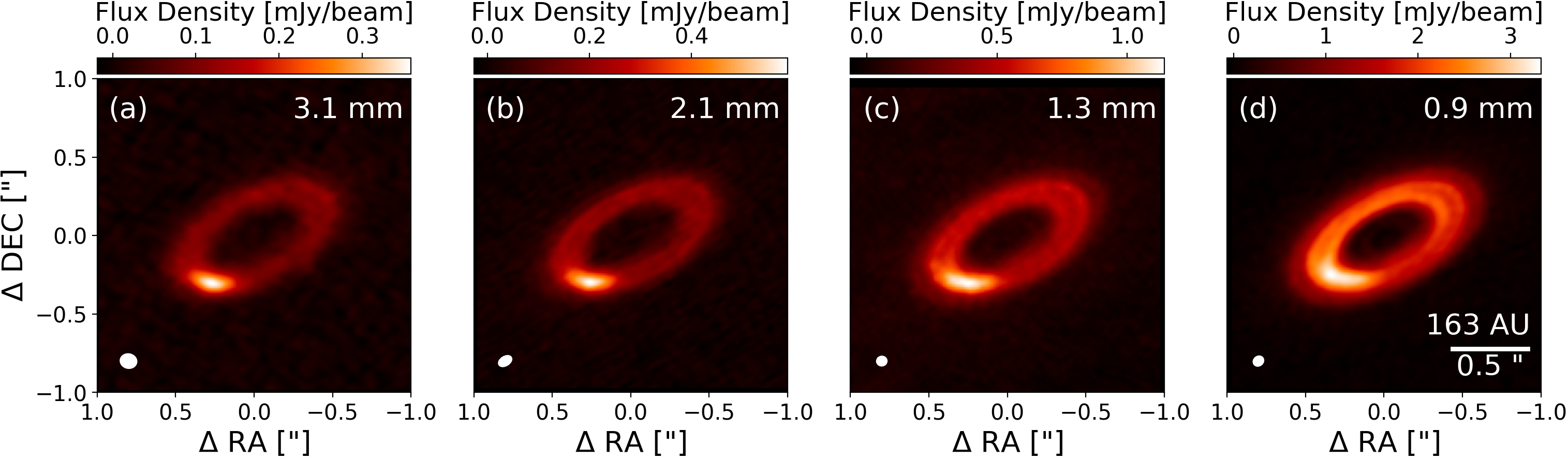}
\caption{ALMA continuum images at 3.1 mm (a), 2.1 mm (b), 1.3 mm (c), and 0.9 mm (d). The synthesized beam size (peak SNR) is 0.095 $\times$ 0.081$''$ (60), 0.081 $\times$ 0.049$''$ (80),  0.056 $\times$ 0.051$''$ (55), and 0.059 $\times$ 0.050$''$ (133) from panel (a) to (d). The fits files are available in the online article.}
\label{fig:images}
\end{center}
\end{figure*}

\subsection{Data Reduction}\label{sec:data_reduction}
Bands 3 and 4 observations were calibrated with the ALMA pipeline in CASA 6.4.1.12 \citep{CASA_2022}, and Band 7 observations with the ALMA pipeline in CASA 6.5.4.9. The data reduction followed the standard procedures adopted by the DSHARP collaboration \citep{Andrews_2018}. For each execution block, channels containing any line emissions were first flagged from the data when present, and the data was averaged in time to 6 seconds and in frequency over 8 channels.

Self-calibration was then performed in CASA version 6.5.1.23, beginning with the SB data. The SB executions were then concatenated with LB executions and self-calibrated together. All execution blocks were aligned in the \textit{u}-\textit{v} plane to the phase center of the LB execution with highest SNR using the exoALMA visibility-alignment tools \citep{loomis_2025}. To boost the signal-to-noise ratio (SNR), we combined scans, spectral windows, and polarizations when computing gain solutions using {\fontfamily{lmtt}\selectfont gaincal}. The images are reconstructed using {\fontfamily{lmtt}\selectfont tclean} with a threshold of 3$\sigma$ and a cell size of tenth the synthesized beam. We employed the Multi-term (Multi-Scale) Multi-Frequency Synthesis deconvolver ({\fontfamily{lmtt}\selectfont mtmfs}) \citep{Rau_2011} with deconvolution scales of [0, 1, 2, 4, 8] beam sizes and {\fontfamily{lmtt}\selectfont nterms = 2}. We adopted {\fontfamily{lmtt}\selectfont briggs} weighting with {\fontfamily{lmtt}\selectfont robust = 0.5}. 

For Band 3 data, three rounds of phase-only self-calibration were executed with progressively shorter solution intervals ({\fontfamily{lmtt}\selectfont solint} = inf, 300 sec, 100 sec), followed by one round of amplitude calibration with {\fontfamily{lmtt}\selectfont solint} = inf. The peak SNR was improved by 42$\%$ after the self-calibration. A final phase self-calibration with {\fontfamily{lmtt}\selectfont solint} = inf was applied to the concatenated dataset, also combining different scans, spectral windows and polarizations. We did not apply amplitude self-calibration because it did not significantly improve the SNR. The final image has SNR $\sim$60 and a synthesized beam of 0.095$\times$ 0.081$''$.

For the Band 4 data, four rounds of phase-only self-calibration were performed with solution intervals of {\fontfamily{lmtt}\selectfont solint} = inf, 300 sec, 100 sec, and 60 sec, followed by one round of amplitude self-calibration with {\fontfamily{lmtt}\selectfont solint} = inf. The peak SNR improved by 60$\%$ after the self-calibration. We did not perform self-calibration on the final concatenated Band 4 dataset, as it did not significantly improve the SNR and instead reduced the effective angular resolution. The final image has SNR $\sim$80 and a synthesized beam of 0.081$\times$ 0.049$''$.

For the Band 7 data, five rounds of phase-only self-calibration were performed with solution intervals of {\fontfamily{lmtt}\selectfont solint} = inf, 300 sec, 100 sec, 60 sec and 20 sec. We did not apply amplitude self-calibration because it improved SNR by $<$5$\%$. The peak SNR improved by 168$\%$ after the self-calibration. Since self-calibration yielded negligible SNR gains and required extensive flagging, we did not apply it to the concatenated dataset. The final image has SNR $\sim$133 and a synthesized beam of 0.059$\times$ 0.050$''$. 

\section{Results}\label{sec:result}

In this section, we present the continuum images across four ALMA bands (\S\ref{sec:image}), and characterize the morphology of the arc across wavelength with visibility modeling (\S\ref{sec:frank}).

\subsection{Images}\label{sec:image}

The reconstructed images at all four wavelengths are shown in Figure~\ref{fig:images}. Rayleigh-Jeans brightness temperature maps are presented in Figure~\ref{fig:image_T}, shown both at the native resolution and convolved to a common beam matched to the largest synthesized beam among the four wavelengths ($0.095 \times 0.081''$). At 3.1 and 2.1 mm, the emission reveals two rings with a bright arc on the southeastern side of the outer ring. At 1.3 and 0.9 mm, the higher angular resolution observation resolves the disk into three rings, with a bright arc situated between the two narrower rings. The background rings and gaps show the characteristic ``double-gap, triple-ring'' structures previously seen in the outer disk of a number of systems, such as TW Hya, HD 169142, AS 209, DoAr 25, Elias 2-20, and in RU Lup \citep{ALMA_2015, Andrews_2016, Perez_2019, Huang_2018}. Such features may indicate the presence of a super-Earth in the middle ring \citep{Dong_2017, Dong_2018_2, Bae_2017}.

HD~34282 had previously been imaged at lower resolution (0.17$\times0.10''$) at 0.9 mm \citep{van_der_Plas_2017}. They modeled the emission with constant-surface-brightness rings on top of the faint broad background ring and a 2D Gaussian arc centered at $r_{\rm arc}\simeq0.43''$ and $\theta_{\rm arc}\simeq18^\circ$. The azimuthal angle $\theta$ is defined in the deprojected disk plane, measured from the redshifted side of the disk major axis increasing counter-clockwisely. We define $\theta_{\rm arc}$ as the azimuthal location of the arc center. Higher-resolution imaging at 0.9 mm (0.067$\times0.054''$) from exoALMA \citep{Curone_2025}, with a resolution comparable to our data, resolves three rings and recovers an arc located at $r_{\rm arc}\simeq0.45''$ and $\theta_{\rm arc}\simeq20^\circ$, consistent with the result of \citet{van_der_Plas_2017}. 

\begin{figure}
\begin{center}
\includegraphics[width=0.48\textwidth]{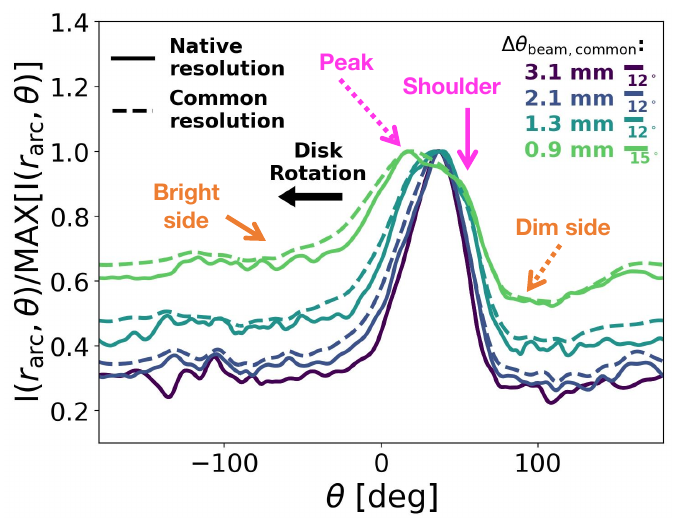}
\caption{Normalized azimuthal brightness profiles at $r_\mathrm{arc}$ (Table~\ref{table:bestfit}) after deprojection using the geometry in Appendix~\ref{app:galario}. The disk-plane azimuth $\theta$ is measured from the red-shifted major axis  and increases counterclockwise. Solid curves show profiles from the native-resolution images; dashed curves are from images convolved to a common $0.095 \times 0.081''$ beam (Band 3). The azimuthal resolution, $\Delta\theta_{\rm beam,common}$, is defined as the tangential FWHM of the deprojected common beam at the arc peak. The black arrow marks clockwise disk rotation \citep{de_Boer_2021}. The 0.9 mm profile shows a sharp peak with a leading-side shoulder (pink arrows)  and asymmetric baselines (orange arrows).}
\label{fig:prof}
\end{center}
\end{figure}

Figure~\ref{fig:prof} shows the deprojected azimuthal intensity profiles evaluated at the arc radius $r_{\rm arc}$ at each wavelength listed in Table~\ref{table:bestfit}, as obtained from parametric $\{u,v\}$-plane fitting with \texttt{galario} \citep{Tazzari_2018} (see Appendix~\ref{app:galario}). The profiles are extracted using the disk inclination,
position angle, and center offsets obtained from the \texttt{galario} fit, and reveal pronounced wavelength-dependent morphology. We convolve all images to a common synthesized beam ($0.095 \times 0.081''$; Band 3 beam; see the lower panel of Figure~\ref{fig:image_T}) to enable a fair comparison of the profiles (dashed lines in Figure~\ref{fig:prof}) across wavelengths. The arc-to-ring contrast increases with wavelength in these common beam profiles: contrast $=$ 1.5, 2.0, 2.8 and 3.5 at 0.9, 1.3, 2.1 and 3.1 mm, respectively, similar to PDS 70 \citep{Doi_2024}. The arc also becomes azimuthally narrower at longer wavelengths. The arc peak shifts from 0.9 to 1.3 mm, but remains consistent at longer wavelengths. Quantitative measurements of the arc azimuthal location and width will be reported in \S\ref{sec:frank} and discussed in \S\ref{sec:vortex_evidence}.

The arc has a different shape at 0.9 mm compared to longer wavelengths. The azimuthal profile in Figure~\ref{fig:prof} exhibits a shallow  shoulder (pink solid arrow) on the leading side
of the arc relative to the disk rotation followed by a sharper peak (pink dashed arrow), whereas the longer-wavelength profiles are closer to a Gaussian shape. The portion of the background ring on the near-side (southwest; orange dashed arrow) is dimmer than the far-side (northeast; orange solid arrow) at both 1.3 and 0.9 mm, and is most pronounced at 0.9 mm. 

\subsection{Measurement of the Arc Morphology}\label{sec:frank}
To facilitate comparison with theoretical predictions of dust trapping (\S\ref{sec:vortex_evidence}), we measure the azimuthal position and width of the arc at each wavelength. Due to the complex arc morphology, it is difficult to define a single parametric model that fits the data across all bands (Figure~\ref{fig:res}; Appendix~\ref{app:galario}), so we instead determine the arc peak and width directly from its azimuthal brightness profile. 

Since the arc lies on top of bright rings, its peak and width can be biased by the underlying axisymmetric emission. To minimize this contamination, we follow a procedure similar to \citet{Curone_2025} to separate the axisymmetric disk emission from the localized arc. The workflow has three steps: 
\begin{enumerate}
    \item We fit a parametric \texttt{galario} visibility model to determine the disk geometry (PA, inclination $i$), phase-center offsets ($\Delta\alpha,\Delta\delta$), and the arc radius ($r_{\rm arc}$) at each wavelength. The visibilities are modeled with three Gaussian rings and a 2D Gaussian arc. Further details are provided in Appendix~\ref{app:galario}. 

    \item We deproject the visibilities using the parameters from step 1 (listed in Table~\ref{table:bestfit}) and use \texttt{frank} \citep{Jennings_2020} to reconstruct a non-parametric axisymmetric radial intensity profile in logarithmic space on a radial grid with $N = 400$ points. The hyper-parameters are set to $\alpha = 1.3$ and $w_\mathrm{smooth} = 0.01$, where $\alpha$ defines the effective signal-to-noise threshold beyond which the data are excluded from the fit, and $w_\mathrm{smooth}$ suppresses noisy oscillations. We set $R_\mathrm{max}$ =3$''$ to encompass all possible extended emission, such that \texttt{frank} assumes zero emission beyond $R_\mathrm{max}$. The resulting \texttt{frank} model is sampled at the same $(u,v)$ coordinates as the observations to generate synthetic model visibilities. Residual visibilities are obtained by subtracting these synthetic visibilities from the observed ones and imaged using the CASA {\fontfamily{lmtt}\selectfont tclean} algorithm.
    \item We subtract the \texttt{frank} model from the visibilities, image the residuals, and measure the arc peak location and full width at half maximum (FWHM) from the deprojected residual profile evaluated at $r_{\rm arc}$.
\end{enumerate}

\begin{figure*}
\begin{center}
\includegraphics[width=\textwidth]{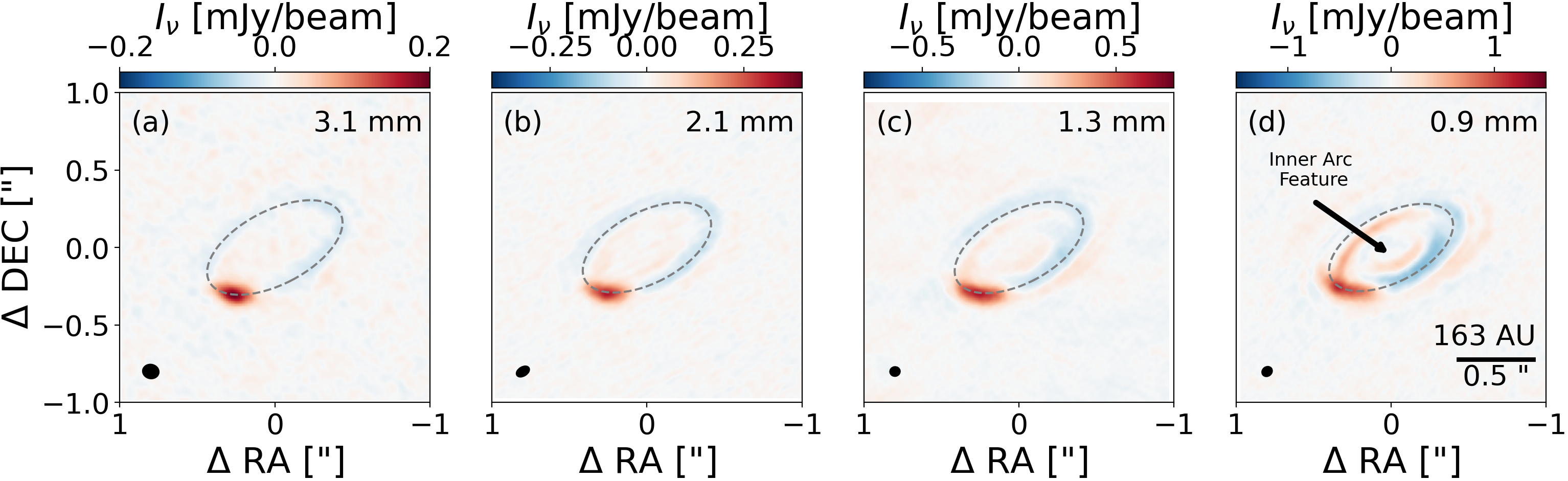}
\caption{Residual map generated by subtracting the \texttt{frank} model from the data at 3.1 mm (a), 2.1 mm (b), 1.3 mm (c), and 0.9 mm (d). The synthesized beam is indicated by the ellipse in the lower left corner. The arc radius $r_{\rm arc}$ is indicated by the dashed gray line. At 0.9 mm, a faint arc-like positive residual is visible at $r \sim 0.17''$ (panel d; arrow).}
\label{fig:res_frank}
\end{center}
\end{figure*}

Figure~\ref{fig:res_frank} shows the \texttt{frank} residual maps at 3.1, 2.1, 1.3, and 0.9 mm. At all wavelengths, a bright non-axisymmetric residual is detected on the southeast side of the disk, tracing the localized arc, with peak SNR of 40, 45, 27, and 43 at 3.1, 2.1, 1.3, and 0.9 mm, respectively. There are also low-level, nearly axisymmetric negative residuals associated with the arc located at $r_{\rm arc}$. Since the \texttt{frank} model only includes the azimuthally averaged (axisymmetric) component, the flux of the localized arc is redistributed evenly around the ring in the model. Subtracting this model therefore produces a positive residual at the arc location and a small negative residual along the rest of the ring, which reduces the peak SNR of the arc.

\begin{figure*}
\begin{center}
\includegraphics[width=\textwidth]{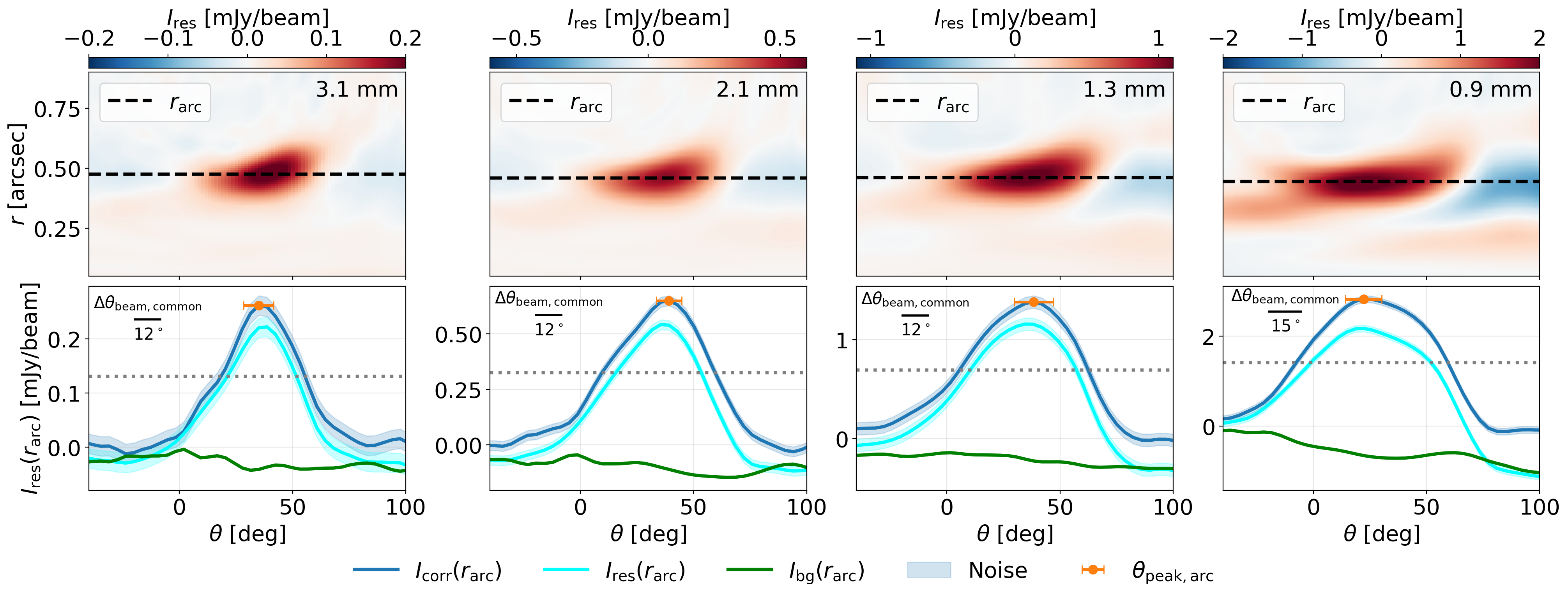}
\caption{Top: Residual maps (Figure~\ref{fig:res_frank}) convolved to a common Band~3 beam ($0.095 \times 0.081''$) and deprojected for comparison across 3.1, 2.1, 1.3, and 0.9 mm (left to right). The horizontal dashed line marks $r_{\rm arc}$. Bottom: Azimuthal profiles of the top panels at $r_{\rm arc}$ (Table~\ref{table:bestfit}). Cyan shows the raw residual $I_{\rm res}$, green is the opposite-sector background $I_{\rm bg}$, and blue is the background-corrected profile $I_{\rm corr}$; the shaded region indicates $\pm3\sigma$ noise. The orange marker marks the peak $\theta_{\rm peak,arc}$ with $3\sigma$ uncertainty. The horizontal dotted line indicates the half-maximum level in $I_{\rm corr}$, and the black bar (top left) the effective azimuthal beam $\Delta\theta_{\mathrm{beam,common}}$ at the arc peak. The arc is fully resolved in the azimuthal direction at all wavelengths.}
\label{fig:res_frank_prof}
\end{center}
\end{figure*}

At 0.9 and 1.3 mm, additional positive and negative residuals appear in the narrow rings at $r\sim r_\mathrm{arc}$, indicating that the rings are dimmer on the near side and brighter on the far side. This asymmetry within the ring is also evident in the deprojected azimuthal brightness profiles (Figure~\ref{fig:prof}), where the portion of the ring on near-side (southwest) is fainter than the far-side (northeast) at both wavelengths indicated by the orange arrows on the 0.9 mm curve. In addition, an extended arc-like positive residual is detected on the near side of the inner disk at 0.9 mm (arrow in Figure~\ref{fig:res_frank}d). These features are unlikely to affect our analysis of the arc and are discussed further in \S\ref{sec:discussion}.

\begin{table}
\centering
\caption{Measured arc peak azimuth and azimuthal width (FWHM). Quoted uncertainties are $1\sigma$ values estimated from bootstrap resampling.}
\label{table:arc_peak_fwhm}
\begin{tabular}{lcc}
\hline\hline
Wavelength & $\theta_{\rm peak,arc}$ [$^\circ$] & $\Delta\theta_{\rm FWHM,arc}$ [$^\circ$] \\
\hline
3.1 mm & $35.0 \pm 2.2$ & $37.1 \pm 1.4$ \\
2.1 mm & $39.2 \pm 1.9$ & $50.0 \pm 0.7$ \\
1.3 mm & $38.3 \pm 2.9$ & $57.0 \pm 0.9$ \\
0.9 mm & $23.1 \pm 2.6$ & $65.8 \pm 0.7$ \\
\hline
\end{tabular}
\end{table}

 To ensure that differences in beam size do not bias the comparison between bands, we convolve all residual images to a common $0.095\times0.081^{\prime\prime}$ beam (the Band~3 beam). We then deproject each residual image into polar coordinates, $I_{\rm res}(r,\theta)$, as shown in the top row of Figure~\ref{fig:res_frank_prof}. At each wavelength, we extract an azimuthal profile at the arc radius $r_{\rm arc}$ (Table~\ref{table:bestfit}; black dashed line in Figure~\ref{fig:res_frank_prof}), shown as the cyan curve in the lower panels of Figure~\ref{fig:res_frank_prof}.

\begin{figure}
\begin{center}
\includegraphics[width=0.45\textwidth]{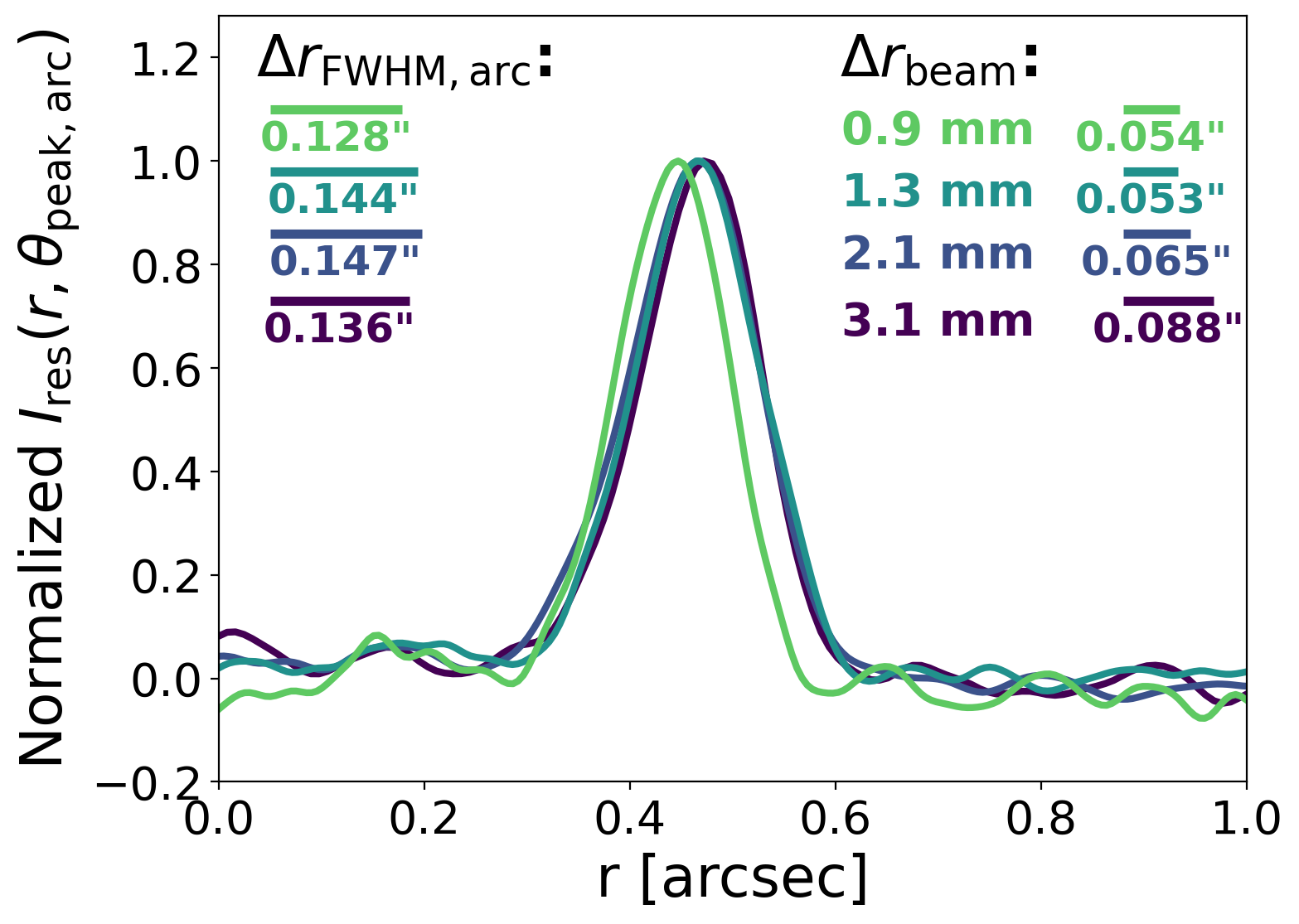}
\caption{Radial profiles of the normalized residual intensity, $I_{\mathrm{res}}(r,\theta_{\mathrm{peak,arc}})$, extracted at the azimuthal peak of the arc for all four wavelengths, measured from the native-resolution images. The effective beam FWHM at each wavelength, $\Delta r_{\mathrm{beam}}$, is listed in the upper right. The FWHM of the arc, $\Delta r_\mathrm{FWHM,arc}$, is listed in the upper left corner. The arc is only marginally resolved in the radial direction at 3.1 mm ($\Delta r_\mathrm{FWHM,arc}=1.5\Delta r_{\mathrm{beam}}$), but is better resolved at shorter wavelengths ($\Delta r_\mathrm{FWHM,arc}>2\Delta r_{\mathrm{beam}}$).}
\label{fig:radial_frank_res}
\end{center}
\end{figure}

To isolate the arc emission, we estimate a local background directly from the azimuthal residual profile at the arc radius. We use the profile from the arc-free side, corresponding to the region located $180^\circ$ away from the arc peak as a reference background (green line in Figure~\ref{fig:res_frank_prof}), under the assumption that it samples the same underlying axisymmetric emission without contribution from the arc. We subtract this background profile point-by-point from the arc-side residual profile, thereby correcting for the artificial negative baseline introduced when the localized arc biases the \texttt{frank} reconstruction. The resulting baseline-corrected profile 
(blue line in Figure~\ref{fig:res_frank_prof}) is then used to measure the azimuthal location and width of the arc.

We estimate the statistical uncertainties of the arc peak location and azimuthal width using a bootstrap approach for 500 realizations. In each realization, we perturb the residual azimuthal profile $I_{\mathrm{res}}(r_\mathrm{arc},\theta)$ by adding Gaussian noise with a standard deviation equal to the measured map RMS listed in Table~\ref{table:observation}, re-estimate and subtract the baseline profiles, and then re-measure the peak location $\theta_{\mathrm{peak,arc}}$ and azimuthal width $\Delta\theta_{\mathrm{FWHM,arc}}$ from the noise-perturbed profile. The uncertainties are taken to be the standard deviations of the resulting bootstrap distributions of $\theta_{\rm peak,arc}$ and $\Delta\theta_{\rm FWHM,arc}$.

\begin{figure*}
\begin{center}
\includegraphics[width=\textwidth]{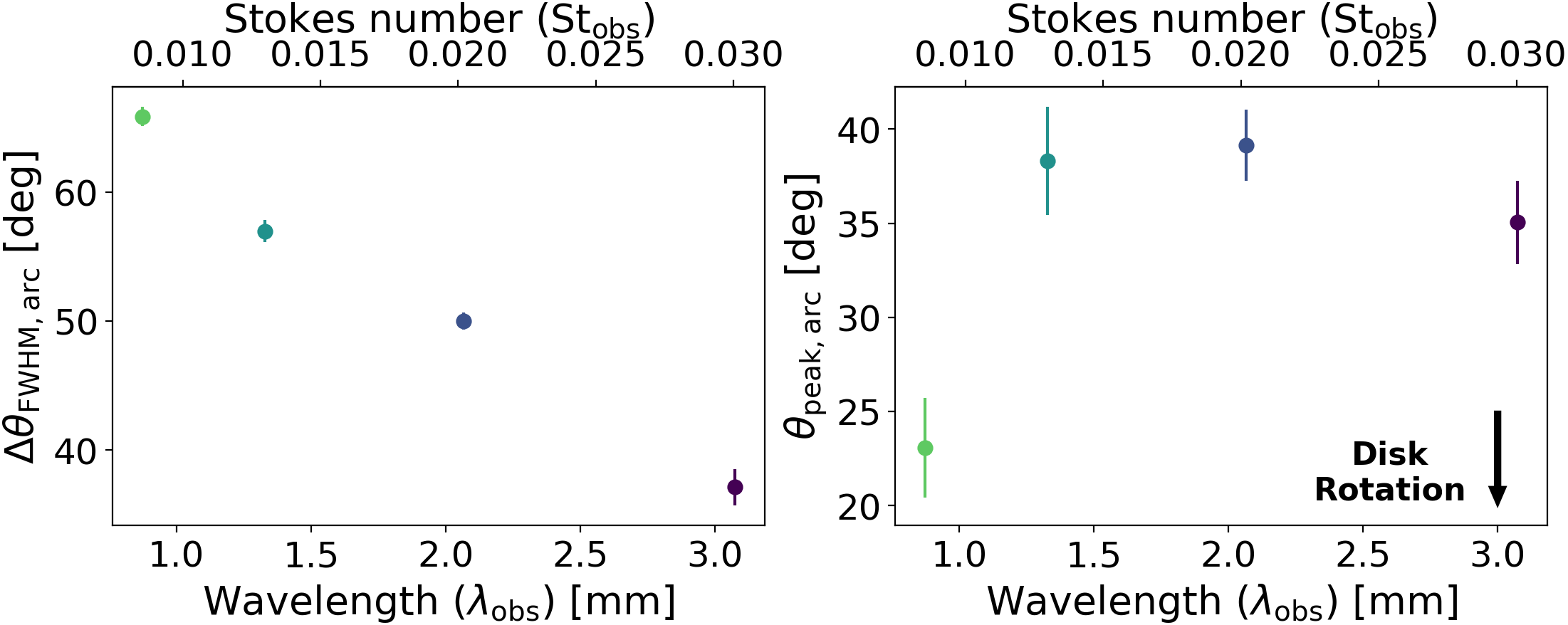}
\caption{Azimuthal full width at half maximum of the arc, $\Delta\theta_{\mathrm{FWHM,arc}}$ (left), and azimuthal peak location, $\theta_{\mathrm{peak,arc}}$ (right), measured from the baseline-corrected azimuthal profiles of the \texttt{frank} residuals ($I_{\rm corr}$; bottom row of Figure~\ref{fig:res_frank_prof}). The measurements are listed in Table~\ref{table:arc_peak_fwhm}. The four points correspond to $\lambda_{\mathrm{obs}}=0.9,\ 1.3,\ 2.1,$ and $3.1$~mm, with observed Stokes numbers $\mathrm{St}_\mathrm{obs}=0.008,\ 0.012,\ 0.020,$ and $0.030$, respectively. Error bars indicate the statistical uncertainties estimated via bootstrap resampling.}
\label{fig:trend}
\end{center}
\end{figure*}

The resulting arc locations and widths are listed in Table~\ref{table:arc_peak_fwhm} and plotted in Figure~\ref{fig:trend}. The measured FWHM decreases monotonically from $\sim65.8\pm0.7^\circ$ at 0.9 mm to $\sim37.1\pm1.4^\circ$ at 3.1 mm. The azimuthal peak shifts by $\sim15\pm4^\circ$ between 0.9 and 1.3 mm, while the relative shifts among the three longer wavelengths remain undetected ($\lesssim4^\circ$) within $\sim1~\sigma_{\theta_{\mathrm{peak},\mathrm{arc}}}$.

The measured azimuthal widths in the common-beam images range from $37^\circ$ to $66^\circ$ from 3.1~mm to 0.9~mm, more than three times larger than the effective azimuthal beam sizes of $12^\circ$ to $15^\circ$ (Figure~\ref{fig:prof}), indicating that the arc is azimuthally resolved.

The deprojected radial profiles of the residual image at the peak azimuth of $I_\mathrm{res}(r_\mathrm{arc},\theta)$, are shown in Figure~\ref{fig:radial_frank_res}.
At 0.9, 1.3, and 2.1 mm, the radial FWHM of the arc is $>2\times$ the beam, while at 3.1 mm, it is only $1.5\times$ the beam. Since the arc is azimuthally well resolved, the marginal radial resolution at 3.1 mm primarily averages the emission over a narrow radial range and should not significantly bias $\theta_{\rm peak,arc}$ or $\Delta\theta_{\rm FWHM,arc}$, unless the azimuthal profile changes rapidly with radius on scales comparable to the beam.

\section{Multi-wavelength Analysis}\label{sec:analysis}

In this section, we estimate the optical depth (\S\ref{sec:optical_depth}) and spectral index (\S\ref{sec:spectral_index}) of the observations, and compare the observations with theoretical predictions to assess the consistency of the observed features with a dust-trapping vortex model (\S\ref{sec:vortex_evidence}).

\subsection{Optical Depth} \label{sec:optical_depth}

\begin{figure}
\begin{center}
\includegraphics[width=0.45\textwidth]{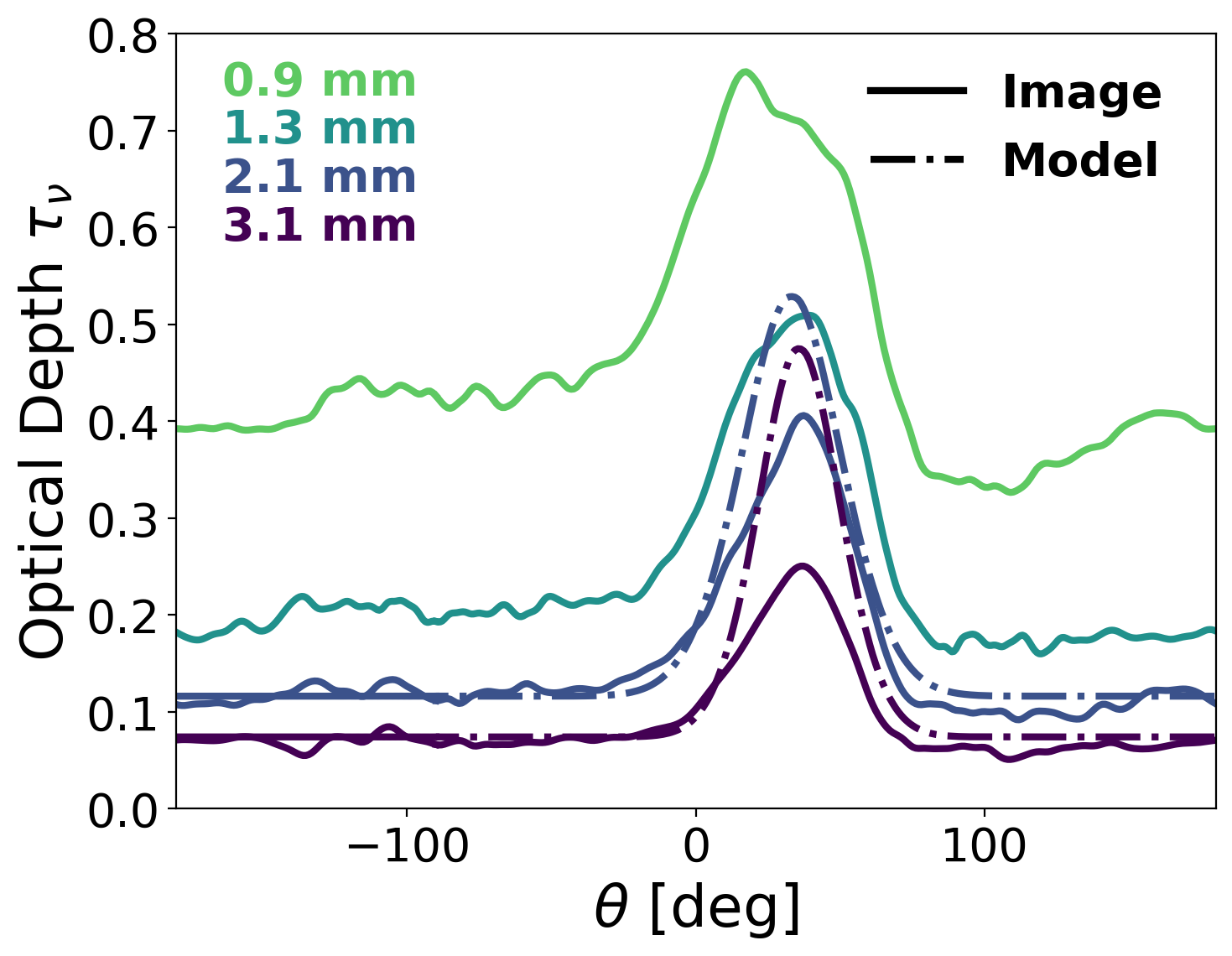}
\caption{Azimuthal optical depth profiles across the arc. Solid curves show $\tau_\nu$ measured directly from the images at all four wavelengths, and dot-dashed curves show $\tau_\nu$ from the best-fit \texttt{galario} parametric models at 3.1 and 2.1 mm, as only the 3.1 and 2.1 mm models reproduce the data well (Figure~\ref{fig:res}).}
\label{fig:tau}
\end{center}
\end{figure}

The observations reveal changes in the arc morphology with wavelength. The observed morphology traces the dust density distribution in the midplane only if the emission is optically thin; otherwise, the emission becomes also sensitive to the temperature distribution. We therefore first estimate the optical depth at each wavelength to determine whether the continuum reliably probes the dust distribution before testing dust-trapping models. We consider emission to be optically thin for optical depth $\tau_\nu\lesssim$ 0.5, marginally optically thick for $\tau_\nu \sim 1$, and optically thick for $\tau_\nu\gtrsim$ 2. 

We estimate $\tau_\nu$ at each wavelength using a toy model based on the Eddington-Barbier approximation in an isothermal slab \citep{Birnstiel_2018} assuming local thermodynamic equilibrium. If we only consider absorption opacity, the optical depth is

\begin{equation}
    \tau_\nu = -\ln\!\left[1 - \frac{I_\nu}{B_\nu(T_{\rm dust})}\right].
    \label{eq:tau}
\end{equation}
where $I_\nu$ is the observed surface brightness at frequency $\nu$, and $T_\mathrm{dust}$ is the dust temperature. We adopt $T_{\rm dust}$ as the midplane temperature from the 2D thermal structure inferred from multiple molecular tracers \citep{Galloway_Sprietsma_2025},
assuming the dust and gas are thermally coupled in the disk midplane. 
At the arc radius, $T_{\rm dust} = 29.6 \pm 0.1$~K. The quoted error reflects only propagated statistical uncertainties and excludes systematics (e.g., emitting-height/vertical-structure assumptions). For the purpose of classifying the emission from the arc as optically thin versus thick, a moderate temperature offset does not change our qualitative conclusions. For example, a $\pm 5$~K change in $T_{\rm dust}$ would alter $\tau_\nu$ by $\sim$25-30\% in the arc.
As azimuthal temperature variations potentially induced by a vortex are expected to be small \citep{Temmink_2025}, we adopt a constant dust temperature with azimuth. 

The azimuthal profiles of the absorption optical depth are shown in Figure~\ref{fig:tau}. The solid curves show $\tau_\nu$ estimated directly from the images. At 3.1, 2.1, 1.3, and 0.9 mm, we measure $\tau_\nu \approx 0.08,\ 0.10,\ 0.19,$ and $0.40$ in the ring, and $\tau_\nu \approx 0.25,\ 0.41,\ 0.50,$ and $0.77$ in the arc, respectively.

The beam dilution makes the unresolved or marginally resolved features appear fainter than their intrinsic brightness, leading to an underestimate of $\tau_\nu$. This effect is expected to be the most significant in Band~3, where the beam is the coarsest and the intrinsic dust concentration is likely the most compact. Although the arc is azimuthally resolved, it is only marginally resolved radially at Band 3, as shown in Figure~\ref{fig:radial_frank_res}. As a result, the image-based estimates of $\tau_\nu$ likely represent lower limits on the intrinsic optical depth of the arc. 

To test this effect, we use the best-fit \texttt{galario} models (Appendix~\ref{app:galario}), which are fitted directly in the visibility plane and therefore account for beam convolution, to carry out an additional optical depth estimate at 3.1 and 2.1 mm. The results are shown as dot-dashed curves in Figure~\ref{fig:tau}. We restrict this analysis to 3.1 and 2.1 mm because only at these wavelengths does the \texttt{galario} model reproduce the data well (Figure~\ref{fig:res}).

At 3.1 mm, the model yields $\tau_\nu \approx 0.47$ at the arc peak, which is higher than $\tau_\nu \approx 0.25$ from the image, consistent with the effect of beam dilution. The ring has $\tau_\nu \approx 0.08$ in both cases. At 2.1 mm, the model gives $\tau_\nu \approx 0.53$ at the arc peak, again higher than the image-based estimate ($\tau_\nu \approx 0.41$), while the ring has $\tau_\nu \approx 0.12$. Even though the arc is radially resolved at 2.1 mm (Figure~\ref{fig:radial_frank_res}), beam convolution and blending with the underlying ring still suppress the observed peak intensity, so the image-based $\tau_\nu$ underestimates the intrinsic value recovered by \texttt{galario}. We therefore adopt the model-based optical depths as fiducial estimates at Band 3 and 4. These values indicate that the arc remains intrinsically optically thin in both bands under a pure-absorption assumption even after accounting for beam dilution, while the surrounding ring is more optically thin. However, until higher-resolution observations are available, we cannot formally exclude the possibility that the arc consists of unresolved optically thick clumps with a beam filling factor smaller than unity, which could reproduce the observed optically thin, beam-averaged emission \citep{Dullemond_2018}.

At the other extreme, an upper limit on the absorption optical depth at 3.1 and 2.1 mm can be derived by assuming the Band 7 emission is optically thick at the arc location, and adopting the corresponding brightness temperature $T_b$ (computed using the full Planck function) as the mid-plane physical dust temperature. Under this assumption, $\tau_\nu \sim 0.45$ at 3.1\,mm and $\tau_\nu \sim 0.82$ at 2.1\,mm, in which the emission is still (marginally) optically thin.

Dust scattering at millimeter wavelengths suppresses the emergent intensity for a given extinction optical depth \citep{Zhu_2019, Liu_2019}. We estimate optical depths including scattering for a given single-scattering albedo $\omega_\nu$ by inverting the relation between $\chi_\nu \equiv I_\nu/B_\nu$ and $\tau_\nu$ (Equation~11 in \cite{Zhu_2019}). We adopt $\chi_\nu \simeq 0.378$ at 3.1~mm and $\chi_\nu \simeq 0.411$ at 2.1~mm for the arc, measured from the best-fit \texttt{galario} models. While $\omega_\nu$ remains poorly constrained in HD 34282, larger $\omega_\nu$ implies a stronger suppression of the emergent intensity and therefore requires a larger $\tau_\nu$ to reproduce the observed $\chi_\nu$. Observational constraints generally suggest that disk albedos can reach $\omega_\nu \lesssim 0.8$ \citep{Ueda_2020,Yoshida_2025}. For this upper end of plausible albedos, we infer $\tau_\nu \lesssim 1.42$ at 3.1 mm and $\tau_\nu \lesssim 1.67$ at 2.1 mm for the arc, and $\tau_\nu \lesssim 0.19$ at 3.1 mm and $\tau_\nu \lesssim 0.30$ at 2.1 mm for the ring. For a representative albedo of $\omega_\nu = 0.7$, the ring remains optically thin, while the arc is (marginally) optically thin at the two wavelengths.

At shorter wavelengths, the similar peak Rayleigh-Jeans brightness temperatures at the arc in Bands~6 ($9.7 \pm 0.2$ K) and 7 ($11.6 \pm 0.1$ K) (Figure~\ref{fig:image_T}) suggest emission approaching optically thick at Band 7. The derived optical depth ($\tau_\nu \sim 0.77$) may therefore be underestimated if dust scattering reduces the observed intensity, implying that the intrinsic optical depth could be higher.

\subsection{Spectral Index Distributions}\label{sec:spectral_index}

\begin{figure*}
\begin{center}
\includegraphics[width=\textwidth]{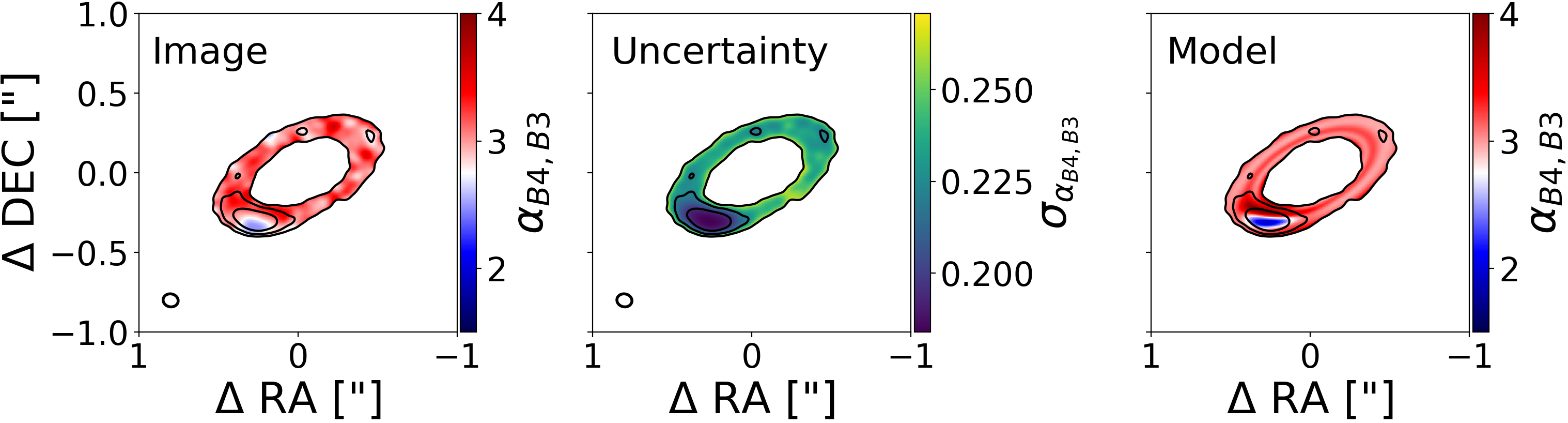}
\caption{Left: Observed spectral index map between ALMA Bands 3 and 4. The common
resolution of $0.095\times0.081''$ (the Band~3 beam) is shown by the ellipse in
the lower-left corner; black contours trace the Band~3 continuum emission.
Middle: Corresponding $1\sigma$ uncertainty map, including contributions from
the image RMS noise and the absolute flux-calibration uncertainties.
Right: Spectral index map between Bands 3 and 4 computed from the best-fit
\texttt{galario} models (Appendix~\ref{app:galario}).}
\label{fig:alpha}
\end{center}
\end{figure*}

We measure the spectral index of the observed intensity as
\begin{equation}\label{eq:spectral_index}
\alpha =
\frac{\ln\!\big(I_{\nu_1}/I_{\nu_2}\big)}
     {\ln(\nu_1/\nu_2)} \, 
\end{equation}
where $I_{\nu_1}$ and $I_{\nu_2}$ are the intensities at frequencies $\nu_1$ and $\nu_2$. We use the data at 3.1 mm and 2.1 mm (Band 3 and 4) to compute $\alpha$, as these bands are most likely to be optically thin (\S\ref{sec:optical_depth}), and more reliable tracers of the dust distribution. Prior to this, we align both images to a common center using the position offsets derived in Appendix~\ref{app:galario}, and convolve them to a common beam as 0.095 $\times$ 0.081$''$ (Band 3 beam). We compute the spectral index only for regions with SNR $>15$ in both bands. The resulting $\alpha_{\mathrm{B4,B3}}$ map is shown in the left panel in Figure~\ref{fig:alpha}.

The spectral index uncertainties include contributions from the image RMS noise and the absolute flux calibration. We adopt a 5$\%$  absolute flux-calibration uncertainty for both bands \citep{ALMA_THB_Cycle12_2025}. The resulting uncertainty map, $\sigma_{\alpha_{\mathrm{B4,B3}}}$, is shown in the middle panel of Figure~\ref{fig:alpha}. The spectral index reaches its minimum at the arc, with $\alpha_{\mathrm{B4,B3}} = 2.48 \pm 0.18$, whereas the surrounding ring at $r_{\rm arc}$ has a higher mean value of $\alpha_{\mathrm{B4,B3}} \simeq 3.11 \pm 0.25$, with a peak value exceeding 3.5.

We also compute the spectral index of the isolated arc emission using the peak intensities of the baseline-corrected profiles ($I_{\rm corr}$; Figure~\ref{fig:res_frank_prof}), obtaining $\alpha_{\mathrm{B4,B3}}^{\rm arc} = 2.29 \pm 0.18$, consistent within uncertainties with the value measured from the observed intensity, indicating that contamination from the underlying ring emission does not significantly affect the spectral index estimate.

Since the arc is not fully resolved radially at 3.1 mm as shown in Figure~\ref{fig:radial_frank_res}, beam smearing can dilute the peak intensity and bias $\alpha$ locally. To assess this effect, we also compute $\alpha$ from the best-fit \texttt{galario} model images as shown in the right panel of Figure~\ref{fig:alpha}. In the model, $\alpha_{\mathrm{B4,B3}}$ reaches values as low as $\sim 1.92$ within the arc, while remaining $\sim 2.98$ in the background.

In vortices, dust is expected to grow rapidly to larger sizes than in the surrounding disk. A lower $\alpha$ at the arc may indicate larger grains compared to the background disk \citep{Birnstiel_2013, Testi_2014,  Birnstiel_2024}, consistent with dust-trapping. 

When dust scattering is included, the interpretation of $\alpha$ becomes degenerate. The peak value of $\alpha \gtrsim 3.5$ in the ring is consistent with optically thin emission from small grains (e.g., $\lesssim 100\,\mu$m). The mean value of $\alpha \sim 3$ across the ring may reflect spatial variations in the dust population or optical depth. Meanwhile, the reduced $\alpha$ observed in the arc can arise either from optically thick emission or from optically thin emission dominated by larger grains \citep{Zhu_2019, Liu_2019}. In the former case, the enhanced optical depth may be due to dust growth to sizes greater than a millimeter, or a concentration of a very large amount of dust, which are not mutually exclusive. Both scenarios imply that the arc hosts a higher dust surface density and/or larger grain sizes than the surrounding background emission. This is consistent with an enhanced dust concentration and grain growth expected in dust-trapping vortices \citep{Birnstiel_2013, Li_2020}.

\subsection{Evidence for Dust Trapping}\label{sec:vortex_evidence}

In this section, we assess the dust-trapping hypothesis for the vortex candidate by comparing the wavelength dependence of the arc with theoretical predictions. The azimuthal width and peak shift are examined in \S\ref{sec:azi_width} and \S\ref{sec:peak_shift}, respectively.

\subsubsection{Azimuthal Width}\label{sec:azi_width}

Both analytical models and numerical simulations predict that larger grains traced by longer-wavelength emission should be more tightly confined azimuthally within a dust-trapping vortex than smaller grains \citep{Birnstiel_2013, Lyra_2013}. This is the result of size-dependent gas-dust coupling. As larger particles are more weakly coupled to the gas, they drift more efficiently toward pressure maxima, resulting in a narrower azimuthal concentration of dust.

Our data follow this prediction. The measured azimuthal widths decrease systematically with wavelength (Figure~\ref{fig:trend}), consistent with size-dependent trapping of larger grains. This trend remains after beam matching and subtraction of the axisymmetric component, indicating that it is intrinsic (Figure~\ref{fig:prof}).

While the observed trend may be influenced by the potentially high optical depth at 0.9 and 1.3 mm, it may be less affected at 2.1 and 3.1 mm, where the emission is expected to be at least marginally optically thin (\S\ref{sec:optical_depth}). In addition, this trend persists when the emission in the background ring is subtracted. The azimuthal FWHM measured from the baseline-corrected \texttt{frank} residual profile (Figure~\ref{fig:trend}) decreases from $\Delta\theta_\mathrm{FWHM,arc}\approx 50.0\pm0.7^\circ$ at 2.1 mm to $ \approx 37.1\pm1.4^\circ$ at 3.1 mm (Table~\ref{table:arc_peak_fwhm}). Since \texttt{frank} removes the axisymmetric component, these widths are minimally affected by the underlying ring and provide a robust estimate of the intrinsic arc width.

To further characterize the morphology of the arc in HD 34282, we estimate its aspect ratio using the observed azimuthal FWHM, $\Delta\theta_{\rm FWHM,arc}$ (Figure~\ref{fig:trend}), and radial FWHM,  $\Delta r_{\rm FWHM,arc}$ (Figure~\ref{fig:radial_frank_res}), at the three shorter wavelengths, where the arc is well resolved in 2D. We find aspect ratios of 2.7, 3.2, and 4.0 at 2.1, 1.3, and 0.9 mm, respectively, indicating a moderately elongated structure. These values are comparable to other arcs, such as the north and south arcs in MWC~758
(3.1 and 4.0, values derived here from Table 1 of \citealt{Dong_2018}). The inferred range is also consistent with the aspect ratios expected for stable vortices ($>$ 2) in analytic models \citep{Lyra_2013}.

\subsubsection{Peak Shift}\label{sec:peak_shift}

In vortices, the azimuthal trapping location of dust particles depends on their Stokes number, $\mathrm{St}$, which quantifies the degree of dust-gas coupling, and can also be affected by vortex self-gravity. For tightly coupled grains ($\mathrm{St} \lesssim 0.1$), dust is trapped close to the pressure maximum and vortex self-gravity is expected to produce negligible azimuthal offsets. For weakly coupled grains, vortex self-gravity can push the dust peak azimuthally ahead of the gas vortex, producing azimuthal shifts in the direction of disk rotation with increasing wavelength \citep{Mittal_2015,Baruteau_2016}. 

We now estimate the Stokes number of the dust probed by our observations, $\mathrm{St}_{\mathrm{obs}}$. In the Epstein regime, 
\begin{equation}
    \mathrm{St}_{\mathrm{obs}} =
    \frac{\pi \rho_s a}{2\,\Sigma_{\mathrm{gas}}(r_{\mathrm{arc}})},
    \label{eq:stokes_num}
\end{equation}
where $a$ is the grain size, $\Sigma_{\mathrm{gas}}$ is the gas surface
density, and $\rho_s$ is the internal grain density. We assume that emission at each wavelength is dominated by a characteristic size $a \sim\lambda_{\mathrm{obs}}/2\pi$ \citep{Draine_2006} and take $\rho_s = 1~\mathrm{g\,cm^{-3}}$. We adopt $\Sigma_{\rm gas}(r_{\rm arc}=0.45'') = 2.6~{\rm g~cm^{-2}}$ inferred from CO kinematics \citep{Longarini_2025}. The Stokes number $\mathrm{St}_{\mathrm{obs}}$ here refers to an observational quantity associated with the grains dominating the emission at a given wavelength \citep{van_der_Marel_2021}, rather than a unique physical Stokes number of the dust population. We obtained $\mathrm{St}_{\mathrm{obs}}$ for $\lambda_{\mathrm{obs}}$ = 3.1, 2.1, 1.3 and 0.9 mm as shown in Figure~\ref{fig:trend}. This yields $\mathrm{St_\mathrm{obs}} \lesssim 0.03$ at all four wavelengths, a regime in which theory predicts negligible azimuthal offsets \citep{Baruteau_2016}.

The observations at $\lambda_{\mathrm{obs}} \geq 1.3$ mm agree with this
prediction. The azimuthal positions of the arc at 1.3, 2.1, and 3.1 mm are
mutually consistent within $\sim$1$\sigma$ (Figure~\ref{fig:trend}). The peak
shifts by only $\sim 1 \pm 3^{\circ}$ between 1.3 and 2.1 mm and by $\sim 4 \pm 3^{\circ}$ between 2.1 and 3.1 mm.

From 0.9 to 1.3~mm, the arc peak shifts by $\sim15 \pm 4^{\circ}$ in the direction opposite to the disk rotation (Figure~\ref{fig:trend}), as inferred from scattered-light imaging and gas kinematics \citep{de_Boer_2021, van_der_Plas_2017}. A qualitatively similar behavior has been reported in HD~135344B, where the arc peak shifts by $\sim50^{\circ}$ between Band~9 and Band~3, also opposite to the direction of disk rotation \citep{Cazzoletti_2018}. Given the very small Stokes numbers in HD 34282, such a large, reversed shift cannot arise from vortex self-gravity and likely reflects other effects, such as optical depth or temperature structure. 

\section{Additional Band 7 Features}\label{sec:discussion}

At Band 7 (0.9 mm), we detect an arc-shoulder feature with a large azimuthal offset from the arc seen at longer wavelengths (pink arrows in Figure~\ref{fig:prof}). The origin of this feature is unclear. One possible explanation is azimuthal temperature variation. If the emission is (marginally) optically thick at Band 7 \citep[e.g., with $\tau_\nu \sim 1.1$,][]{van_der_Plas_2017}, the brightness distribution may partly trace the temperature rather than the dust column density.  Simulations show that gas perturbations in an anticyclonic vortex co-rotate with the vortex and can displace slightly hotter gas outward and cooler gas inward, producing a downstream temperature enhancement \citep{Lovelace_2014}. 

It is not yet clear whether a comparable effect should occur for the dust. Testing it with the dust emission data presented here will require predictions generated using coupled dust-gas radiative transfer modeling.

Images at 0.9 mm exhibit a localized dimming of two narrow rings on the southwest side of the disk. This is reflected in the deprojected azimuthal brightness at the arc radius (Figure~\ref{fig:prof}), where the portion of the ring at near-side of the arc (orange dashed arrow) is fainter than the far-side (orange solid arrow). The origin of this dim feature is unclear. It could reflect azimuthal temperature variations caused by shadows, or changes in optical depth and scattering geometry that reduce the apparent brightness in dust emission. Additional scattered-light observations probing a potentially misaligned inner disk, combined with radiative transfer modeling, might be helpful to clarify the origin of this feature.

We also detect a faint, azimuthally extended arc at $r \sim 0.17''$ with $\sim7\sigma$ significance, appearing as an arc-like positive residual in the \texttt{frank} residual map (arrow in Figure~\ref{fig:res_frank}d). 
This feature is not detected in scattered light data \citep{de_Boer_2021,Quiroz_2022}, but it lies close to the inner working angle, where sensitivity to small-separation structure is limited. Recent non-redundant masking interferometry has revealed a ring-like component at $r\sim0.23''$ in the $L'$ band \citep{Vides_2025}, broadly comparable to the feature identified here. 

\section{Conclusions}\label{sec:conclusions}

We present new ALMA continuum observations of the HD~34282 protoplanetary disk at 0.9, 2.1, and 3.1~mm, together with archival 1.3~mm data, all with sub-$0.1''$ resolution (Figure~\ref{fig:images}). The disk hosts an inner cavity, and the outer disk shows a characteristic compact double-gap, triple-ring morphology (more pronounced at shorter wavelengths), similar to the pattern found in a number of other disks, such as HL Tau and HD 169142. The disk also shows a prominent arc that is azimuthally resolved at all wavelengths (Figure~\ref{fig:prof}). In the radial direction, its FWHM is $>2\times$ the beam size at 0.9, 1.3, and 2.1 mm, and $\approx1.5\times$ the beam size at 3.1 mm (Figure~\ref{fig:radial_frank_res}). We test the dust-trapping interpretation of the arc by comparing its wavelength-dependent morphology with theoretical predictions. Our main conclusions are:

\begin{enumerate}
    \item The azimuthal width of the arc decreases monotonically toward longer wavelengths from 65.8$\pm$0.7$^\circ$ at 0.9 mm to 37.1$\pm$1.4$^\circ$ at 3.1 mm (\S\ref{sec:azi_width}; Figure~\ref{fig:trend}; Table~\ref{table:arc_peak_fwhm}). This trend is consistent with dust trapping \citep{Birnstiel_2013}, in which larger grains traced by longer wavelengths (and thus higher Stokes number, St) are more tightly confined azimuthally.

    \item  The dust in HD 34282 probed by our observations is estimated to have $\mathrm{St_\mathrm{obs}} \lesssim 0.03$. For such tightly coupled dust, vortex models predict negligible azimuthal peak shifts due to vortex self-gravity \citep{Mittal_2015,Baruteau_2016}. Our 1.3, 2.1 and 3.1 mm measurements are consistent with this expectation (\S\ref{sec:peak_shift}; Figure~\ref{fig:trend}). 
    In contrast, the azimuthal peak differs by $15\pm4^\circ$ between 0.9 and 1.3~mm. The peak shifts opposite to the disk rotation as wavelength increases, similar to HD~135344B \citep{Cazzoletti_2018}. This behavior may reflect optical-depth or temperature effects.

    \item The arc structure at 2.1 and 3.1 mm is well reproduced by parametric \texttt{galario} visibility models consisting of radial Gaussian rings and a 2D Gaussian arc (Appendix~\ref{app:galario}). We estimate the optical depth from both the ALMA images and the \texttt{galario} model by comparing the measured brightness temperature with the expected dust temperature. In both cases, the brightness temperature is well below the dust temperature, implying optically thin emission at these wavelengths (\S\ref{sec:optical_depth}; Figure~\ref{fig:tau}). This conclusion remains valid when dust scattering is included for plausible albedos of $\omega_\nu\simeq0.7$. However, the arc may be optically thick at 0.9 mm.
    
    \item The 2.1-3.1 mm spectral index within the arc ($\alpha\sim2.5$) is lower than that in the background rings ($\alpha\sim3$), potentially suggesting moderately optically thick emission and/or optically thin emission dominated by mm-sized grains in the arc, as predicted by theory \citep{Li_2020} (\S\ref{sec:spectral_index}; Figure~\ref{fig:alpha}).

    \item At 0.9~mm we detect additional structure beyond the ring and arc at the arc radius $r_\mathrm{arc}=0.44''$, including an arc-shoulder feature (pink arrows in Figure~\ref{fig:prof}), and a localized dimming of the ring on the southwest side (orange dashed arrow in Figure~\ref{fig:prof}). We also detect a faint, azimuthally extended arc at $r \sim 0.17''$ (arrow in Figure~\ref{fig:res_frank}d). The origins of these structures are unclear, but they may reflect azimuthal temperature variations and/or optical-depth and scattering effects (\S\ref{sec:discussion}).

\end{enumerate}

Overall, our observations suggest that the HD 34282 arc traces dust trapping in a long-lived pressure maximum, potentially a dust-trapping vortex. A definitive test will require longer-wavelength continuum data to constrain the grain-size distribution, together with high-resolution, high-sensitivity molecular-line kinematics to search for local deviations from Keplerian rotation \citep{Huang_2019}.

\section*{Acknowledgements}
We are grateful to an anonymous referee for constructive suggestions that improved our paper. We thank Bin Ren, Eve Lee, Gregory J. Herczeg, Haochang Jiang, John Carpenter, Lile Wang, Milou Temmink, Nagayoshi Ohashi, Shangjia Zhang, and Yangfan Shi for their insightful comments and suggestions, and Logan Francis for providing the processed Band 6 data. This study was supported by the National Key R\&D Program of China (Grant No. 2025YFE0123100). This research was enabled in part by support provided by the High-performance Computing Platform of Peking University, and the Digital Research Alliance of Canada (\url{alliance.can.ca}). This work was also supported by ADVANCED 149943 and 2024-1.2.8-T\'ET-IPARI-CN-2025-00036 grants, which have been implemented with the support provided by the Ministry of Culture and Innovation of Hungary from the National Research, Development and Innovation Fund, financed under the NKKP ADVANCED and 2024-1.2.8-T\'ET-IPARI-CN funding schemes. H.B.L. is supported by the National Science and Technology Council (NSTC) of Taiwan (Grant Nos. 113-2112-M-110-022-MY3).

\appendix
\section{Brightness Temperature Images} \label{app:image_cb}
\begin{figure}[h!]
\begin{center}
\includegraphics[width=\textwidth]{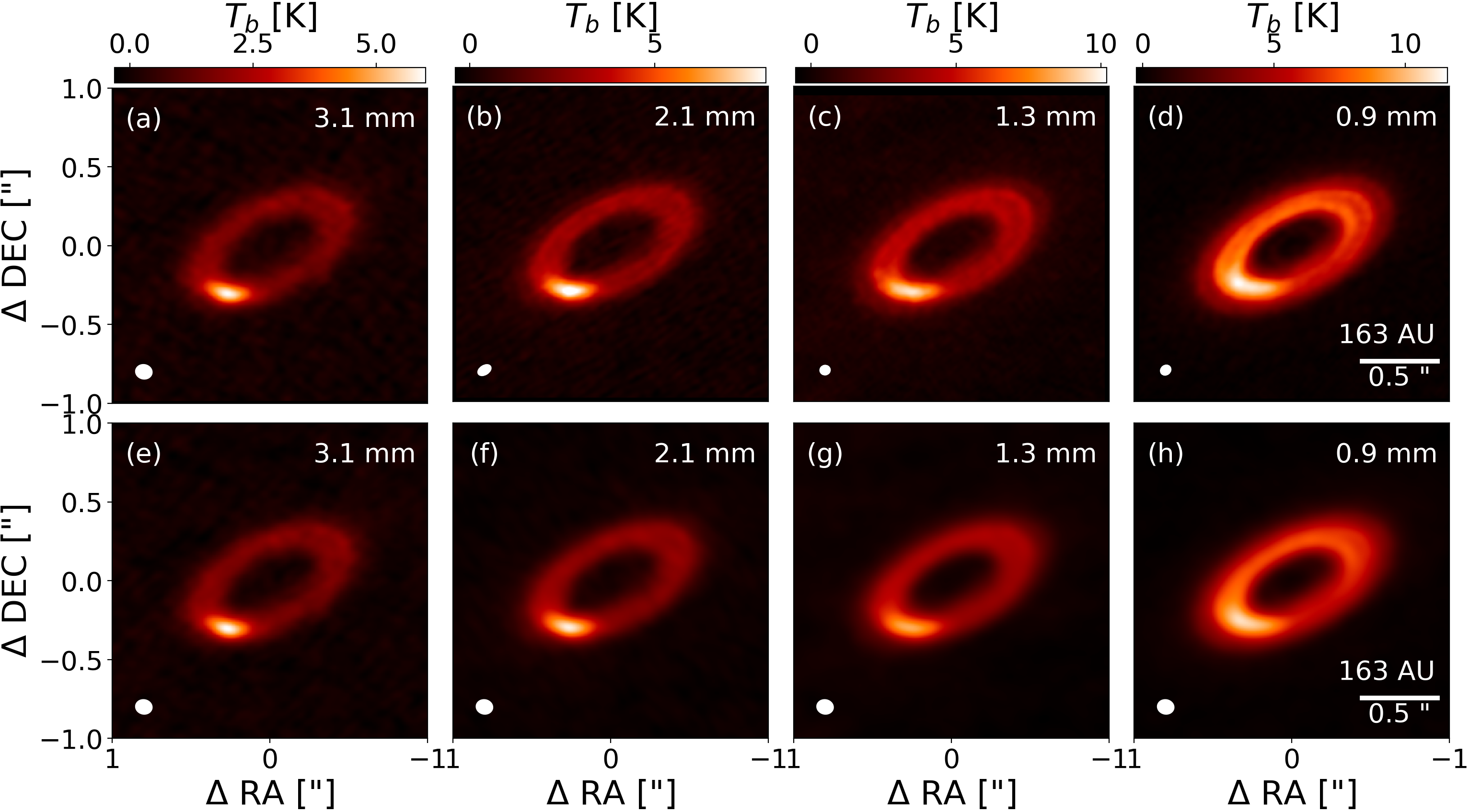}
\caption{Rayleigh-Jeans brightness temperature maps at (a) 3.1 mm, (b) 2.1 mm, (c) 1.3 mm, and (d) 0.9 mm. The synthesized beam sizes (peak S/N) are $0.095 \times 0.081''$ (60), $0.081 \times 0.049''$ (80), $0.056 \times 0.051''$ (55), and $0.059 \times 0.050''$ (133) from panels (a) to (d), respectively. Panels (e) to (h) show the corresponding maps convolved to a common angular resolution, matched to the largest beam among the four bands ($0.095 \times 0.081''$), which is indicated in each panel.}
\label{fig:image_T}
\end{center}
\end{figure}

We show the Rayleigh-Jeans brightness temperature maps in Figure~\ref{fig:image_T}.

\section{Disk Modeling with \texttt{galario}} \label{app:galario}
In this section, we fit parametric \texttt{galario} models to the visibilities at all four wavelengths. For each wavelength, we fit for the position angle ($\mathrm{PA}$), inclination ($i$), and the offsets between the phase center and disk center, $(\Delta\alpha,\Delta\delta)$. We adopt the global morphological model of \citet{Curone_2025} for HD 34282, in which the surface brightness $I_\nu$ in disk-plane polar coordinates $(r,\theta)$ consists of three axisymmetric Gaussian rings and a 2D Gaussian arc:
\begin{equation}
    I(r,\theta) = \sum_{k=1}^{3} f_k \exp\!\left[-\frac{(r-r_k)^2}{2\sigma_k^2}\right] + f_{\rm arc}
    \exp\!\left[-\frac{(r-r_{\rm arc})^2}{2\sigma_{r,{\rm arc}}^2}\right]
    \exp\!\left[-\frac{(\theta-\theta_{\rm arc})^2}{2\sigma_{\theta,{\rm arc}}^2}\right]
    \label{eq:I_model}
\end{equation}
The amplitudes $f_k$ and $f_{\rm arc}$ are sampled uniformly in log space. The model visibilities, $V_{\rm mod}$, are generated by Fourier transforming the sky brightness model after projecting it onto the plane of the sky, and then sampling the result at the observed $(u_j, v_j)$ coordinates:
\begin{equation}
    V_{\rm mod}(u_j,v_j)=\mathcal{F}\{I_\nu(x,y)\}\big|_{(u_j,v_j)}
    \label{eq:vis_model}
\end{equation}

\begin{table*}[htbp]
\centering
\small
\caption{Best-fit results from \texttt{galario} for the three-ring plus one-arc model of HD~34282}
\label{table:bestfit}
\begin{tabular}{lcccc}
\hline\hline
Parameter & B3 (3.1 mm) & B4 (2.1 mm) & B6 (1.3 mm) & B7 (0.9 mm) \\
\hline
\multicolumn{5}{l}{\textit{Ring 1}} \\
$\log_{10}(f_1)$ (Jy sr$^{-1}$) &
$8.006_{-0.022}^{+0.025}$ &
$8.519_{-0.012}^{+0.013}$ &
$8.9687_{-0.0069}^{+0.0058}$ &
$9.6028_{-0.0030}^{+0.0030}$  \\
$r_1$ (mas) &
$358_{-21}^{+19}$ &
$354_{-12}^{+11}$ &
$409_{-2}^{+16}$ &
$384.0_{-3.1}^{+3.9}$ \\
$\sigma_1$ (mas) &
$363_{-16}^{+15}$ &
$373.8_{-9.1}^{+9.5}$ &
$405.4_{-2.8}^{+2.5}$ &
$430.5_{-2.2}^{+2.0}$ \\
\hline
\multicolumn{5}{l}{\textit{Ring 2}} \\
$\log_{10}(f_2)$ (Jy sr$^{-1}$) &
$8.497_{-0.053}^{+0.046}$ &
$8.978_{-0.031}^{+0.027}$ &
$9.385_{-0.015}^{+0.016}$ &
$10.141_{-0.016}^{+0.018}$  \\
$r_2$ (mas) &
$371.9_{-3.3}^{+3.0}$ &
$364.2_{-2.5}^{+2.2}$ &
$347.66_{-0.79}^{+0.86}$ &
$360.1_{-0.8}^{+1.1}$  \\
$\sigma_2$ (mas) &
$35.8_{-4.2}^{+4.0}$ &
$42.9_{-2.2}^{+2.0}$ &
$42.0_{-1.8}^{+1.7}$ &
$59.3_{-1.3}^{+1.9}$ \\
\hline
\multicolumn{5}{l}{\textit{Ring 3}} \\
$\log_{10}(f_3)$ (Jy sr$^{-1}$) &
$8.7059_{-0.0072}^{+0.0068}$ &
$9.2169_{-0.0044}^{+0.0044}$ &
$9.7954_{-0.0025}^{+0.0024}$ &
$10.3207_{-0.0056}^{+0.0043}$  \\
$r_3$ (mas) &
$509.3_{-5.2}^{+4.7}$ &
$505.0_{-3.1}^{+2.8}$ &
$486.7_{-1.7}^{+1.5}$ &
$505.7_{-2.6}^{+3.2}$  \\
$\sigma_3$ (mas) &
$79.7_{-4.6}^{+4.6}$ &
$87.7_{-2.4}^{+2.5}$ &
$108.5_{-1.2}^{+1.1}$ &
$109.6_{-1.6}^{+1.3}$  \\

\hline
\multicolumn{5}{l}{\textit{arc}} \\
$\log_{10}(f_{\rm arc})$ (Jy sr$^{-1}$) &
$9.384_{-0.014}^{+0.014}$ &
$9.7018_{-0.0057}^{+0.0061}$ &
$10.0072_{-0.0036}^{+0.0035}$ &
$10.2565_{-0.0033}^{+0.0034}$  \\
$r_{\rm arc}$ (mas) &
$475.1_{-2.1}^{+2.1}$ &
$459.2_{-1.1}^{+1.0}$ &
$461.28_{-0.64}^{+0.71}$ &
$442.13_{-0.40}^{+0.50}$  \\
$\sigma_{r,{\rm arc}}$ (mas) &
$35.3_{-1.1}^{+1.1}$ &
$44.21_{-0.69}^{+0.62}$ &
$49.14_{-0.46}^{+0.68}$ &
$48.45_{-0.38}^{+0.39}$ \\
$\theta_{\rm arc}$ (deg) &
$35.77_{-0.40}^{+0.32}$ &
$33.05_{-0.18}^{+0.18}$ &
$30.795_{-0.091}^{+0.078}$ &
$13.77_{-0.18}^{+0.16}$  \\
$\sigma_{\theta,{\rm arc}}$ (deg) &
$15.18_{-0.27}^{+0.27}$ &
$18.62_{-0.16}^{+0.16}$ &
$21.273_{-0.067}^{+0.089}$ &
$28.90_{-0.16}^{+0.15}$ \\
\hline
\multicolumn{5}{l}{\textit{Geometry}} \\
Inclination $i$ (deg) &
$60.109_{-0.097}^{+0.093}$ &
$59.381_{-0.044}^{+0.051}$ &
$59.443_{-0.0001}^{+0.0001}$ &
$59.523_{-0.013}^{+0.014}$  \\
PA (deg) &
$117.56_{-0.12}^{+0.12}$ &
$117.40_{-0.06}^{+0.07}$ &
$117.905_{-0.0001}^{+0.0001}$ &
$117.583_{-0.016}^{+0.015}$  \\
$\Delta\alpha$ (mas) &
$10.09_{-0.76}^{+0.76}$ &
$18.91_{-0.35}^{+0.35}$ &
$-28.85_{-0.12}^{+0.10}$ &
$29.30_{-0.09}^{+0.09}$ \\
$\Delta\delta$ (mas) &
$-11.99_{-0.60}^{+0.63}$ &
$-25.10_{-0.28}^{+0.24}$ &
$59.48_{-0.11}^{+0.08}$ &
$-18.06_{-0.07}^{+0.06}$  \\
\hline
\hline
\end{tabular}

\tablefoot{%
\textbf{Note.} Columns correspond to different observing bands (B3, B4, B6 and B7), while rows list the fitted parameters. For the ring and arc components, we report the median of the marginalized posterior distribution and the corresponding statistical uncertainties from the 16th and 84th percentiles of the MCMC marginalized distributions.
The geometry section lists the inclination ($i$), position angle (PA), and positional offsets ($\Delta\alpha$, $\Delta\delta$) for each band.}
\end{table*}

We adopt a standard Gaussian likelihood, equivalent to minimizing the weighted complex $\chi^2$:
\begin{equation}
    \ln\mathcal{L} = -\frac{1}{2}\sum_j w_j\Big[ \left( \operatorname{Re} V_{{\rm data},j} - \operatorname{Re} V_{{\rm mod},j} \right)^2 + \left( \operatorname{Im} V_{{\rm data},j} - \operatorname{Im} V_{{\rm mod},j} \right)^2 \Big]
    \label{eq:likelihood}
\end{equation}
where $V_{{\rm data},j}$ are the observed visibilities and $w_j$ are their corresponding statistical weights. The likelihood is computed directly on the full set of unbinned visibilities. The image pixel scale is chosen to satisfy Nyquist sampling for the observed baselines, and the field of view is set just large enough to enclose the disk emission. This choice preserves the angular resolution across the disk while minimizing computational cost. We sample the posterior distribution of the model parameters (Eq.~\ref{eq:likelihood}) using a Markov Chain Monte Carlo (MCMC) approach via \texttt{emcee} \citep{Foreman-Mackey_2013}. 

To ensure a robust and efficient exploration of the parameter space, we fit the model with a two-step coarse-to-fine MCMC scheme, similar to the procedure used in \cite{Curone_2025}. A first coarse run uses uniformly seeded walkers within physically motivated bounds to locate the main posterior modes. A second fine run reinitialized walkers in a tight Gaussian ball around the coarse-run medians ($\sigma$ = $10^{-4}$ for each parameter in the same units as the corresponding parameter) and resamples with the same likelihood for high-precision posteriors. We start with the high-S/N Band 7 data, then use its posterior medians to initialize independent fits for Bands 3, 4, and 6, allowing band-to-band parameter variations while improving stability and reproducibility.

\begin{figure}
\begin{center}
\includegraphics[width=0.7\textwidth]{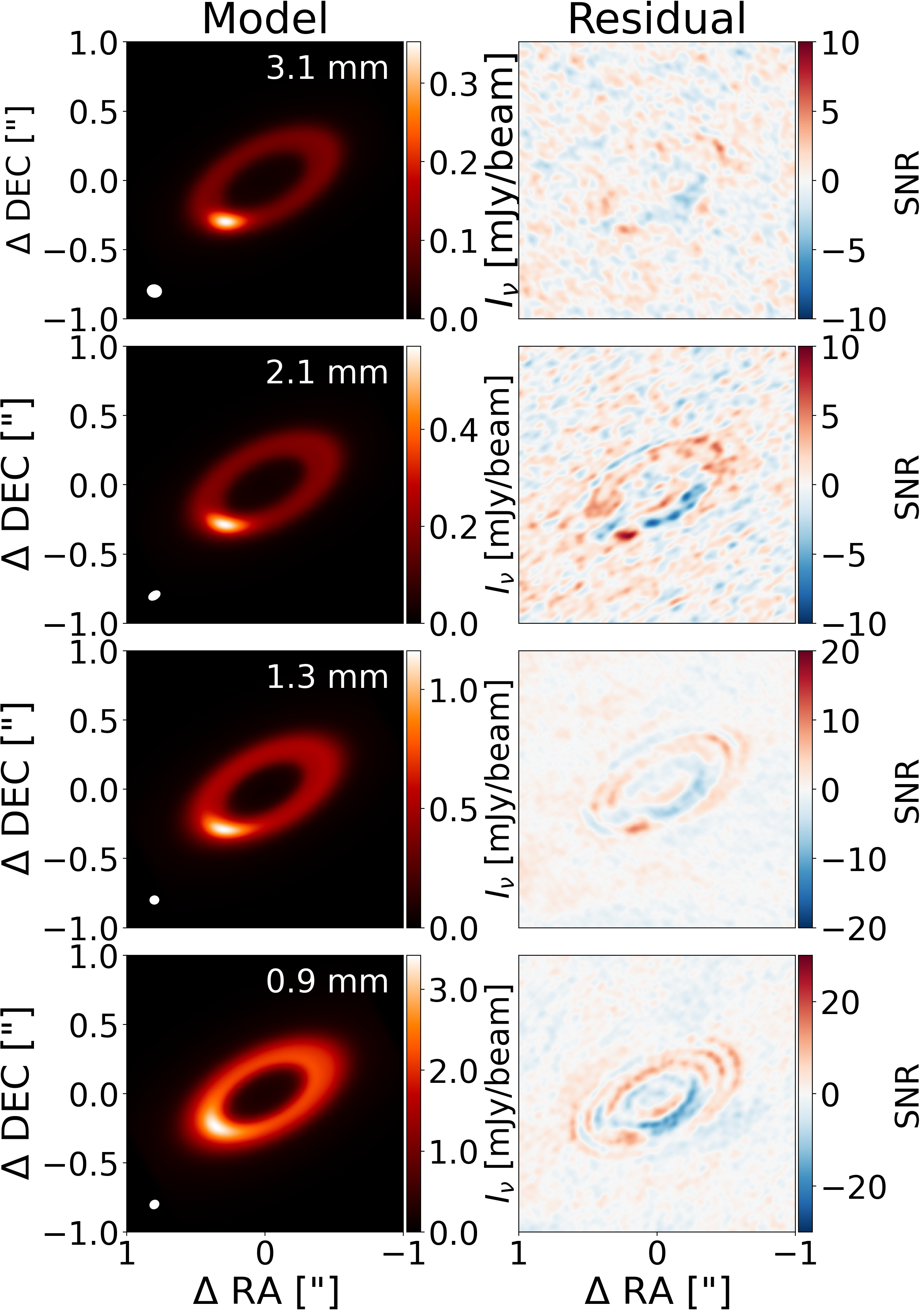}
\caption{Left: models generated with the best-fit visibility parameters (Table \ref{table:bestfit}) at 3.1, 2.1, 1.3, and 0.9 mm (top to bottom). The synthesized beams are shown as white ellipses. Right: corresponding residual maps in signal-to-noise ratio (SNR).}
\label{fig:res}
\end{center}
\end{figure}

The best-fit parameters are listed in Table~\ref{table:bestfit}. Across all wavelengths, we find consistent geometric parameters, with an inclination of $\sim$ 60$^{\circ}$ and a position angle of $\sim$ 117.5$^{\circ}$, in agreement with previous ALMA observations at 0.9 mm \citep{van_der_Plas_2017, Curone_2025}. Figure~\ref{fig:res} presents the synthesized images of the best-fit models and the corresponding residuals. At 3.1 and 2.1 mm, the residuals are low and noise-like, indicating that the ring-arc morphology is well reproduced. At 1.3 and 0.9 mm, residual structures at $>10\sigma$ level remain, suggesting additional substructures not captured by our simple model.

\bibliography{references}

@ARTICLE{Huang_2019,
       author = {{Huang}, Pinghui and {Dong}, Ruobing and {Li}, Hui and {Li}, Shengtai and {Ji}, Jianghui},
        title = "{The Observability of Vortex-driven Spiral Arms in Protoplanetary Disks: Basic Spiral Properties}",
      journal = {\apjl},
         year = 2019,
        month = oct,
       volume = {883},
       number = {2},
          eid = {L39},
        pages = {L39},
          doi = {10.3847/2041-8213/ab40c4},
archivePrefix = {arXiv},
       eprint = {1909.00706},
 primaryClass = {astro-ph.EP},
       adsurl = {https://ui.adsabs.harvard.edu/abs/2019ApJ...883L..39H}
}

@ARTICLE{Huang_2018,
       author = {{Huang}, Jane and {Andrews}, Sean M. and {Dullemond}, Cornelis P. and
        {Isella}, Andrea and {P{\'e}rez}, Laura M. and {Guzm{\'a}n},
        Viviana V. and {{\"O}berg}, Karin I. and {Zhu}, Zhaohuan and
        {Zhang}, Shangjia and {Bai}, Xue-Ning and {Benisty}, Myriam and
        {Birnstiel}, Tilman and {Carpenter}, John M. and {Hughes}, A.
        Meredith and {Ricci}, Luca and {Weaver}, Erik and {Wilner},
        David J.},
        title = "{The Disk Substructures at High Angular Resolution Project (DSHARP). II. Characteristics of Annular Substructures}",
      journal = {\apj},
         year = 2018,
        month = Dec,
       volume = {869},
          eid = {L42},
        pages = {L42},
          doi = {10.3847/2041-8213/aaf740},
archivePrefix = {arXiv},
       eprint = {1812.04041},
 primaryClass = {astro-ph.EP},
       adsurl = {https://ui.adsabs.harvard.edu/\#abs/2018ApJ...869L..42H}
}

@ARTICLE{Tazzari_2018,
       author = {{Tazzari}, Marco and {Beaujean}, Frederik and {Testi}, Leonardo},
        title = "{GALARIO: a GPU accelerated library for analysing radio interferometer observations}",
      journal = {\mnras},
         year = 2018,
        month = jun,
       volume = {476},
       number = {4},
        pages = {4527-4542},
          doi = {10.1093/mnras/sty409},
archivePrefix = {arXiv},
       eprint = {1709.06999},
 primaryClass = {astro-ph.IM},
       adsurl = {https://ui.adsabs.harvard.edu/abs/2018MNRAS.476.4527T}
}

@ARTICLE{Curone_2025,
       author = {{Curone}, Pietro and {Facchini}, Stefano and {Andrews}, Sean M. and {Testi}, Leonardo and {Benisty}, Myriam and {Czekala}, Ian and {Huang}, Jane and {Ilee}, John D. and {Isella}, Andrea and {Lodato}, Giuseppe and {Loomis}, Ryan A. and {Stadler}, Jochen and {Winter}, Andrew J. and {Bae}, Jaehan and {Barraza-Alfaro}, Marcelo and {Cataldi}, Gianni and {Cuello}, Nicol{\'a}s and {Fasano}, Daniele and {Flock}, Mario and {Fukagawa}, Misato and {Galloway-Sprietsma}, Maria and {Garg}, Himanshi and {Hall}, Cassandra and {Izquierdo}, Andr{\'e}s F. and {Kanagawa}, Kazuhiro and {Lesur}, Geoffroy and {Longarini}, Cristiano and {Menard}, Francois and {Orihara}, Ryuta and {Pinte}, Christophe and {Price}, Daniel J. and {Rosotti}, Giovanni and {Teague}, Richard and {Wafflard-Fernandez}, Gaylor and {Wilner}, David J. and {W{\"o}lfer}, Lisa and {Yen}, Hsi-Wei and {Yoshida}, Tomohiro C. and {Zawadzki}, Brianna},
        title = "{exoALMA. IV. Substructures, Asymmetries, and the Faint Outer Disk in Continuum Emission}",
      journal = {\apjl},
         year = 2025,
        month = may,
       volume = {984},
       number = {1},
          eid = {L9},
        pages = {L9},
          doi = {10.3847/2041-8213/adc438},
archivePrefix = {arXiv},
       eprint = {2504.18725},
 primaryClass = {astro-ph.EP},
       adsurl = {https://ui.adsabs.harvard.edu/abs/2025ApJ...984L...9C}
}

@article{Foreman-Mackey_2013,
   author = {{Foreman-Mackey}, Daniel and {Hogg}, David W. and {Lang}, Dustin and {Goodman}, Jonathan},
    title = "{emcee: The MCMC Hammer}",
  journal = {\pasp},
archivePrefix = "arXiv",
   eprint = {1202.3665},
     year = 2013,
    month = mar,
   volume = 125,
   number = 925,
    pages = {306},
      doi = {10.1086/670067},
   adsurl = {https://ui.adsabs.harvard.edu/abs/2013PASP..125..306F}
}

@article{Francis_2020,
   title={Dust-depleted Inner Disks in a Large Sample of Transition Disks through Long-baseline ALMA Observations},
   volume={892},
   ISSN={1538-4357},
   url={http://dx.doi.org/10.3847/1538-4357/ab7b63},
   DOI={10.3847/1538-4357/ab7b63},
   number={2},
   journal={The Astrophysical Journal},
   publisher={American Astronomical Society},
   author={Francis, Logan and van der Marel, Nienke},
   year={2020},
   month=apr, pages={111} }

@misc{loomis_2025,
      title={exoALMA II: Data Calibration and Imaging Pipeline}, 
      author={Ryan A. Loomis and Stefano Facchini and Myriam Benisty and Pietro Curone and John D. Ilee and Gianni Cataldi and Hsi-Wei Yen and Richard Teague and Christophe Pinte and Jane Huang and Himanshi Garg and Ryuta Orihara and Ian Czekala and Brianna Zawadzki and Sean M. Andrews and David J. Wilner and Jaehan Bae and Marcelo Barraza-Alfaro and Daniele Fasano and Mario Flock and Misato Fukagawa and Maria Galloway-Sprietsma and Andres F. Izquierdo and Kazuhiro Kanagawa and Geoffroy Lesur and Cristiano Longarini and Francois Menard and Daniel J. Price and Giovanni Rosotti and Jochen Stadler and Gaylor Wafflard-Fernandez and Lisa Wolfer and Tomohiro C. Yoshida},
      year={2025},
      eprint={2504.19870},
      archivePrefix={arXiv},
      primaryClass={astro-ph.EP},
      url={https://arxiv.org/abs/2504.19870}, 
}

@ARTICLE{CASA_2022,
       author = {{CASA Team} and {Bean}, Ben and {Bhatnagar}, Sanjay and {Castro}, Sandra and {Donovan Meyer}, Jennifer and {Emonts}, Bjorn and {Garcia}, Enrique and {Garwood}, Robert and {Golap}, Kumar and {Gonzalez Villalba}, Justo and {Harris}, Pamela and {Hayashi}, Yohei and {Hoskins}, Josh and {Hsieh}, Mingyu and {Jagannathan}, Preshanth and {Kawasaki}, Wataru and {Keimpema}, Aard and {Kettenis}, Mark and {Lopez}, Jorge and {Marvil}, Joshua and {Masters}, Joseph and {McNichols}, Andrew and {Mehringer}, David and {Miel}, Renaud and {Moellenbrock}, George and {Montesino}, Federico and {Nakazato}, Takeshi and {Ott}, Juergen and {Petry}, Dirk and {Pokorny}, Martin and {Raba}, Ryan and {Rau}, Urvashi and {Schiebel}, Darrell and {Schweighart}, Neal and {Sekhar}, Srikrishna and {Shimada}, Kazuhiko and {Small}, Des and {Steeb}, Jan-Willem and {Sugimoto}, Kanako and {Suoranta}, Ville and {Tsutsumi}, Takahiro and {van Bemmel}, Ilse M. and {Verkouter}, Marjolein and {Wells}, Akeem and {Xiong}, Wei and {Szomoru}, Arpad and {Griffith}, Morgan and {Glendenning}, Brian and {Kern}, Jeff},
        title = "{CASA, the Common Astronomy Software Applications for Radio Astronomy}",
      journal = {\pasp},
         year = 2022,
        month = nov,
       volume = {134},
       number = {1041},
          eid = {114501},
        pages = {114501},
          doi = {10.1088/1538-3873/ac9642},
archivePrefix = {arXiv},
       eprint = {2210.02276},
 primaryClass = {astro-ph.IM},
       adsurl = {https://ui.adsabs.harvard.edu/abs/2022PASP..134k4501C}
}

@article{Andrews_2018,
   title={The Disk Substructures at High Angular Resolution Project (DSHARP). I. Motivation, Sample, Calibration, and Overview},
   volume={869},
   ISSN={2041-8213},
   url={http://dx.doi.org/10.3847/2041-8213/aaf741},
   DOI={10.3847/2041-8213/aaf741},
   number={2},
   journal={The Astrophysical Journal Letters},
   publisher={American Astronomical Society},
   author={Andrews, Sean M. and Huang, Jane and Perez, Laura M. and Isella, Andrea and Dullemond, Cornelis P. and Kurtovic, Nicolás T. and Guzmán, Viviana V. and Carpenter, John M. and Wilner, David J. and Zhang, Shangjia and Zhu, Zhaohuan and Birnstiel, Tilman and Bai, Xue-Ning and Benisty, Myriam and Hughes, A. Meredith and Öberg, Karin I. and Ricci, Luca},
   year={2018},
   month=dec, pages={L41} }

@article{Rau_2011,
   title={A multi-scale multi-frequency deconvolution algorithm for synthesis imaging in radio interferometry},
   volume={532},
   ISSN={1432-0746},
   url={http://dx.doi.org/10.1051/0004-6361/201117104},
   DOI={10.1051/0004-6361/201117104},
   journal={Astronomy \&amp; Astrophysics},
   publisher={EDP Sciences},
   author={Rau, U. and Cornwell, T. J.},
   year={2011},
   month=jul, pages={A71} }

@article{Long_2018,
   title={Gaps and Rings in an ALMA Survey of Disks in the Taurus Star-forming Region},
   volume={869},
   ISSN={1538-4357},
   url={http://dx.doi.org/10.3847/1538-4357/aae8e1},
   DOI={10.3847/1538-4357/aae8e1},
   number={1},
   journal={The Astrophysical Journal},
   publisher={American Astronomical Society},
   author={Long, Feng and Pinilla, Paola and Herczeg, Gregory J. and Harsono, Daniel and Dipierro, Giovanni and Pascucci, Ilaria and Hendler, Nathan and Tazzari, Marco and Ragusa, Enrico and Salyk, Colette and Edwards, Suzan and Lodato, Giuseppe and van de Plas, Gerrit and Johnstone, Doug and Liu, Yao and Boehler, Yann and Cabrit, Sylvie and Manara, Carlo F. and Menard, Francois and Mulders, Gijs D. and Nisini, Brunella and Fischer, William J. and Rigliaco, Elisabetta and Banzatti, Andrea and Avenhaus, Henning and Gully-Santiago, Michael},
   year={2018},
   month=dec, pages={17} }

@ARTICLE{van_der_Marel_2021,
       author = {{van der Marel}, Nienke and {Birnstiel}, Til and {Garufi}, Antonio and {Ragusa}, Enrico and {Christiaens}, Valentin and {Price}, Daniel J. and {Sallum}, Steph and {Muley}, Dhruv and {Francis}, Logan and {Dong}, Ruobing},
        title = "{On the Diversity of Asymmetries in Gapped Protoplanetary Disks}",
      journal = {\aj},
         year = 2021,
        month = jan,
       volume = {161},
       number = {1},
          eid = {33},
        pages = {33},
          doi = {10.3847/1538-3881/abc3ba},
archivePrefix = {arXiv},
       eprint = {2010.10568},
 primaryClass = {astro-ph.EP},
       adsurl = {https://ui.adsabs.harvard.edu/abs/2021AJ....161...33V}
}

@ARTICLE{van_der_Marel_2016,
       author = {{van der Marel}, N. and {Cazzoletti}, P. and {Pinilla}, P. and {Garufi}, A.},
        title = "{Vortices and Spirals in the HD135344B Transition Disk}",
      journal = {\apj},
         year = 2016,
        month = dec,
       volume = {832},
       number = {2},
          eid = {178},
        pages = {178},
          doi = {10.3847/0004-637X/832/2/178},
archivePrefix = {arXiv},
       eprint = {1607.05775},
 primaryClass = {astro-ph.EP},
       adsurl = {https://ui.adsabs.harvard.edu/abs/2016ApJ...832..178V}
}

@ARTICLE{Zhu_2014b,
       author = {{Zhu}, Zhaohuan and {Stone}, James M.},
        title = "{Dust Trapping by Vortices in Transitional Disks: Evidence for Non-ideal Magnetohydrodynamic Effects in Protoplanetary Disks}",
      journal = {\apj},
         year = 2014,
        month = nov,
       volume = {795},
       number = {1},
          eid = {53},
        pages = {53},
          doi = {10.1088/0004-637X/795/1/53},
archivePrefix = {arXiv},
       eprint = {1405.2790},
 primaryClass = {astro-ph.SR},
       adsurl = {https://ui.adsabs.harvard.edu/abs/2014ApJ...795...53Z}
}

@ARTICLE{Regaly_2012,
       author = {{Reg{\'a}ly}, Zs. and {Juh{\'a}sz}, A. and {S{\'a}ndor}, Zs. and {Dullemond}, C.~P.},
        title = "{Possible planet-forming regions on submillimetre images}",
      journal = {\mnras},
         year = 2012,
        month = jan,
       volume = {419},
       number = {2},
        pages = {1701-1712},
          doi = {10.1111/j.1365-2966.2011.19834.x},
archivePrefix = {arXiv},
       eprint = {1109.6177},
 primaryClass = {astro-ph.SR},
       adsurl = {https://ui.adsabs.harvard.edu/abs/2012MNRAS.419.1701R}
}

@ARTICLE{Flock_2015,
       author = {{Flock}, M. and {Ruge}, J.~P. and {Dzyurkevich}, N. and {Henning}, Th. and {Klahr}, H. and {Wolf}, S.},
        title = "{Gaps, rings, and non-axisymmetric structures in protoplanetary disks. From simulations to ALMA observations}",
      journal = {\aap},
         year = 2015,
        month = feb,
       volume = {574},
          eid = {A68},
        pages = {A68},
          doi = {10.1051/0004-6361/201424693},
archivePrefix = {arXiv},
       eprint = {1411.2736},
 primaryClass = {astro-ph.EP},
       adsurl = {https://ui.adsabs.harvard.edu/abs/2015A&A...574A..68F}
}

@ARTICLE{Ataiee_2013,
       author = {{Ataiee}, S. and {Pinilla}, P. and {Zsom}, A. and {Dullemond}, C.~P. and {Dominik}, C. and {Ghanbari}, J.},
        title = "{Asymmetric transition disks: Vorticity or eccentricity?}",
      journal = {\aap},
         year = 2013,
        month = may,
       volume = {553},
          eid = {L3},
        pages = {L3},
          doi = {10.1051/0004-6361/201321125},
archivePrefix = {arXiv},
       eprint = {1304.1736},
 primaryClass = {astro-ph.EP},
       adsurl = {https://ui.adsabs.harvard.edu/abs/2013A&A...553L...3A}
}

@article{Baruteau_2016,
   title={Gas and dust hydrodynamical simulations of massive lopsided transition discs -- II. Dust concentration},
   volume={458},
   ISSN={1365-2966},
   url={http://dx.doi.org/10.1093/mnras/stv2527},
   DOI={10.1093/mnras/stv2527},
   number={4},
   journal={Monthly Notices of the Royal Astronomical Society},
   publisher={Oxford University Press (OUP)},
   author={Baruteau, Clement and Zhu, Zhaohuan},
   year={2016},
   month=apr, pages={3927--3941} }

@ARTICLE{Gaia_2023,
       author = {{Gaia Collaboration} and {Vallenari}, A. and {Brown}, A.~G.~A. and {Prusti}, T. and {de Bruijne}, J.~H.~J. and {Arenou}, F. and {Babusiaux}, C. and {Biermann}, M. and {Creevey}, O.~L. and {Ducourant}, C. and {Evans}, D.~W. and {Eyer}, L. and {Guerra}, R. and {Hutton}, A. and {Jordi}, C. and {Klioner}, S.~A. and {Lammers}, U.~L. and {Lindegren}, L. and {Luri}, X. and {Mignard}, F. and {Panem}, C. and {Pourbaix}, D. and {Randich}, S. and {Sartoretti}, P. and {Soubiran}, C. and {Tanga}, P. and {Walton}, N.~A. and {Bailer-Jones}, C.~A.~L. and {Bastian}, U. and {Drimmel}, R. and {Jansen}, F. and {Katz}, D. and {Lattanzi}, M.~G. and {van Leeuwen}, F. and {Bakker}, J. and {Cacciari}, C. and {Casta{\~n}eda}, J. and {De Angeli}, F. and {Fabricius}, C. and {Fouesneau}, M. and {Fr{\'e}mat}, Y. and {Galluccio}, L. and {Guerrier}, A. and {Heiter}, U. and {Masana}, E. and {Messineo}, R. and {Mowlavi}, N. and {Nicolas}, C. and {Nienartowicz}, K. and {Pailler}, F. and {Panuzzo}, P. and {Riclet}, F. and {Roux}, W. and {Seabroke}, G.~M. and {Sordo}, R. and {Th{\'e}venin}, F. and {Gracia-Abril}, G. and {Portell}, J. and {Teyssier}, D. and {Altmann}, M. and {Andrae}, R. and {Audard}, M. and {Bellas-Velidis}, I. and {Benson}, K. and {Berthier}, J. and {Blomme}, R. and {Burgess}, P.~W. and {Busonero}, D. and {Busso}, G. and {C{\'a}novas}, H. and {Carry}, B. and {Cellino}, A. and {Cheek}, N. and {Clementini}, G. and {Damerdji}, Y. and {Davidson}, M. and {de Teodoro}, P. and {Nu{\~n}ez Campos}, M. and {Delchambre}, L. and {Dell'Oro}, A. and {Esquej}, P. and {Fern{\'a}ndez-Hern{\'a}ndez}, J. and {Fraile}, E. and {Garabato}, D. and {Garc{\'\i}a-Lario}, P. and {Gosset}, E. and {Haigron}, R. and {Halbwachs}, J. -L. and {Hambly}, N.~C. and {Harrison}, D.~L. and {Hern{\'a}ndez}, J. and {Hestroffer}, D. and {Hodgkin}, S.~T. and {Holl}, B. and {Jan{\ss}en}, K. and {Jevardat de Fombelle}, G. and {Jordan}, S. and {Krone-Martins}, A. and {Lanzafame}, A.~C. and {L{\"o}ffler}, W. and {Marchal}, O. and {Marrese}, P.~M. and {Moitinho}, A. and {Muinonen}, K. and {Osborne}, P. and {Pancino}, E. and {Pauwels}, T. and {Recio-Blanco}, A. and {Reyl{\'e}}, C. and {Riello}, M. and {Rimoldini}, L. and {Roegiers}, T. and {Rybizki}, J. and {Sarro}, L.~M. and {Siopis}, C. and {Smith}, M. and {Sozzetti}, A. and {Utrilla}, E. and {van Leeuwen}, M. and {Abbas}, U. and {{\'A}brah{\'a}m}, P. and {Abreu Aramburu}, A. and {Aerts}, C. and {Aguado}, J.~J. and {Ajaj}, M. and {Aldea-Montero}, F. and {Altavilla}, G. and {{\'A}lvarez}, M.~A. and {Alves}, J. and {Anders}, F. and {Anderson}, R.~I. and {Anglada Varela}, E. and {Antoja}, T. and {Baines}, D. and {Baker}, S.~G. and {Balaguer-N{\'u}{\~n}ez}, L. and {Balbinot}, E. and {Balog}, Z. and {Barache}, C. and {Barbato}, D. and {Barros}, M. and {Barstow}, M.~A. and {Bartolom{\'e}}, S. and {Bassilana}, J. -L. and {Bauchet}, N. and {Becciani}, U. and {Bellazzini}, M. and {Berihuete}, A. and {Bernet}, M. and {Bertone}, S. and {Bianchi}, L. and {Binnenfeld}, A. and {Blanco-Cuaresma}, S. and {Blazere}, A. and {Boch}, T. and {Bombrun}, A. and {Bossini}, D. and {Bouquillon}, S. and {Bragaglia}, A. and {Bramante}, L. and {Breedt}, E. and {Bressan}, A. and {Brouillet}, N. and {Brugaletta}, E. and {Bucciarelli}, B. and {Burlacu}, A. and {Butkevich}, A.~G. and {Buzzi}, R. and {Caffau}, E. and {Cancelliere}, R. and {Cantat-Gaudin}, T. and {Carballo}, R. and {Carlucci}, T. and {Carnerero}, M.~I. and {Carrasco}, J.~M. and {Casamiquela}, L. and {Castellani}, M. and {Castro-Ginard}, A. and {Chaoul}, L. and {Charlot}, P. and {Chemin}, L. and {Chiaramida}, V. and {Chiavassa}, A. and {Chornay}, N. and {Comoretto}, G. and {Contursi}, G. and {Cooper}, W.~J. and {Cornez}, T. and {Cowell}, S. and {Crifo}, F. and {Cropper}, M. and {Crosta}, M. and {Crowley}, C. and {Dafonte}, C. and {Dapergolas}, A. and {David}, M. and {David}, P. and {de Laverny}, P. and {De Luise}, F. and {De March}, R.},
        title = "{Gaia Data Release 3. Summary of the content and survey properties}",
      journal = {\aap},
         year = 2023,
        month = jun,
       volume = {674},
          eid = {A1},
        pages = {A1},
          doi = {10.1051/0004-6361/202243940},
archivePrefix = {arXiv},
       eprint = {2208.00211},
 primaryClass = {astro-ph.GA},
       adsurl = {https://ui.adsabs.harvard.edu/abs/2023A&A...674A...1G}
}

@ARTICLE{Birnstiel_2013,
       author = {{Birnstiel}, T. and {Dullemond}, C.~P. and {Pinilla}, P.},
        title = "{Lopsided dust rings in transition disks}",
      journal = {\aap},
         year = 2013,
        month = feb,
       volume = {550},
          eid = {L8},
        pages = {L8},
          doi = {10.1051/0004-6361/201220847},
archivePrefix = {arXiv},
       eprint = {1301.1976},
 primaryClass = {astro-ph.EP},
       adsurl = {https://ui.adsabs.harvard.edu/abs/2013A&A...550L...8B}
}

@article{van_der_Plas_2017,
   title={An 80 au cavity in the disk around HD 34282},
   volume={607},
   ISSN={1432-0746},
   url={http://dx.doi.org/10.1051/0004-6361/201731392},
   DOI={10.1051/0004-6361/201731392},
   journal={Astronomy \&amp; Astrophysics},
   publisher={EDP Sciences},
   author={van der Plas, G. and Menard, F. and Canovas, H. and Avenhaus, H. and Casassus, S. and Pinte, C. and Caceres, C. and Cieza, L.},
   year={2017},
   month=nov, pages={A55} }

@article{Stolker_2016,
   title={Shadows cast on the transition disk of HD 135344B: Multiwavelength VLT/SPHERE polarimetric differential imaging⋆},
   volume={595},
   ISSN={1432-0746},
   url={http://dx.doi.org/10.1051/0004-6361/201528039},
   DOI={10.1051/0004-6361/201528039},
   journal={Astronomy \&amp; Astrophysics},
   publisher={EDP Sciences},
   author={Stolker, T. and Dominik, C. and Avenhaus, H. and Min, M. and de Boer, J. and Ginski, C. and Schmid, H. M. and Juhasz, A. and Bazzon, A. and Waters, L. B. F. M. and Garufi, A. and Augereau, J.-C. and Benisty, M. and Boccaletti, A. and Henning, Th. and Langlois, M. and Maire, A.-L. and Menard, F. and Meyer, M. R. and Pinte, C. and Quanz, S. P. and Thalmann, C. and Beuzit, J.-L. and Carbillet, M. and Costille, A. and Dohlen, K. and Feldt, M. and Gisler, D. and Mouillet, D. and Pavlov, A. and Perret, D. and Petit, C. and Pragt, J. and Rochat, S. and Roelfsema, R. and Salasnich, B. and Soenke, C. and Wildi, F.},
   year={2016},
   month=nov, pages={A113} }

@article{Boehler_2018,
   title={The Complex Morphology of the Young Disk MWC 758: Spirals and Dust Clumps around a Large Cavity},
   volume={853},
   ISSN={1538-4357},
   url={http://dx.doi.org/10.3847/1538-4357/aaa19c},
   DOI={10.3847/1538-4357/aaa19c},
   number={2},
   journal={The Astrophysical Journal},
   publisher={American Astronomical Society},
   author={Boehler, Y. and Ricci, L. and Weaver, E. and Isella, A. and Benisty, M. and Carpenter, J. and Grady, C. and Shen, Bo-Ting and Tang, Ya-Wen and Perez, L.},
   year={2018},
   month=feb, pages={162} }

@ARTICLE{Dong_2018,
       author = {{Dong}, Ruobing and {Liu}, Sheng-yuan and {Eisner}, Josh and {Andrews}, Sean and {Fung}, Jeffrey and {Zhu}, Zhaohuan and {Chiang}, Eugene and {Hashimoto}, Jun and {Liu}, Hauyu Baobab and {Casassus}, Simon and {Esposito}, Thomas and {Hasegawa}, Yasuhiro and {Muto}, Takayuki and {Pavlyuchenkov}, Yaroslav and {Wilner}, David and {Akiyama}, Eiji and {Tamura}, Motohide and {Wisniewski}, John},
        title = "{The Eccentric Cavity, Triple Rings, Two-armed Spirals, and Double Clumps of the MWC 758 Disk}",
      journal = {\apj},
         year = 2018,
        month = jun,
       volume = {860},
       number = {2},
          eid = {124},
        pages = {124},
          doi = {10.3847/1538-4357/aac6cb},
archivePrefix = {arXiv},
       eprint = {1805.12141},
 primaryClass = {astro-ph.SR},
       adsurl = {https://ui.adsabs.harvard.edu/abs/2018ApJ...860..124D}
}

@ARTICLE{Ueda_2020,
       author = {{Ueda}, Takahiro and {Kataoka}, Akimasa and {Tsukagoshi}, Takashi},
        title = "{Scattering-induced Intensity Reduction: Large Mass Content with Small Grains in the Inner Region of the TW Hya disk}",
      journal = {\apj},
         year = 2020,
        month = apr,
       volume = {893},
       number = {2},
          eid = {125},
        pages = {125},
          doi = {10.3847/1538-4357/ab8223},
archivePrefix = {arXiv},
       eprint = {2003.09353},
 primaryClass = {astro-ph.EP},
       adsurl = {https://ui.adsabs.harvard.edu/abs/2020ApJ...893..125U}
}

@ARTICLE{Boehler_2021,
       author = {{Boehler}, Y. and {M{\'e}nard}, F. and {Robert}, C.~M.~T. and {Isella}, A. and {Pinte}, C. and {Gonzalez}, J.-F. and {van der Plas}, G. and {Weaver}, E. and {Teague}, R. and {Garg}, H. and {M{\'e}heut}, H.},
        title = "{Vortex-like kinematic signal, spirals, and beam smearing effect in the HD 142527 disk}",
      journal = {\aap},
         year = 2021,
        month = jun,
       volume = {650},
          eid = {A59},
        pages = {A59},
          doi = {10.1051/0004-6361/202040089},
archivePrefix = {arXiv},
       eprint = {2103.13474},
 primaryClass = {astro-ph.EP},
       adsurl = {https://ui.adsabs.harvard.edu/abs/2021A&A...650A..59B}
}

@article{Kraus_2017,
   title={Dust-trapping Vortices and a Potentially Planet-triggered Spiral Wake in the Pre-transitional Disk of V1247 Orionis},
   volume={848},
   ISSN={2041-8213},
   url={http://dx.doi.org/10.3847/2041-8213/aa8edc},
   DOI={10.3847/2041-8213/aa8edc},
   number={1},
   journal={The Astrophysical Journal Letters},
   publisher={American Astronomical Society},
   author={Kraus, Stefan and Kreplin, Alexander and Fukugawa, Misato and Muto, Takayuki and Sitko, Michael L. and Young, Alison K. and Bate, Matthew R. and Grady, Carol and Harries, Tim T. and Monnier, John D. and Willson, Matthew and Wisniewski, John},
   year={2017},
   month=oct, pages={L11} }

@article{de_Boer_2021,
   title={Possible single-armed spiral in the protoplanetary disk around HD 34282},
   volume={649},
   ISSN={1432-0746},
   url={http://dx.doi.org/10.1051/0004-6361/201936787},
   DOI={10.1051/0004-6361/201936787},
   journal={Astronomy\&amp; Astrophysics},
   publisher={EDP Sciences},
   author={de Boer, J. and Ginski, C. and Chauvin, G. and Menard, F. and Benisty, M. and Dominik, C. and Maaskant, K. and Girard, J. H. and van der Plas, G. and Garufi, A. and Perrot, C. and Stolker, T. and Avenhaus, H. and Bohn, A. and Delboulbe, A. and Jaquet, M. and Buey, T. and Möller-Nilsson, O. and Pragt, J. and Fusco, T.},
   year={2021},
   month=may, pages={A25} }

@ARTICLE{Marr_2022,
       author = {{Marr}, Metea and {Dong}, Ruobing},
        title = "{The Appearance of Vortices in Protoplanetary Disks in Near-infrared Scattered Light}",
      journal = {\apj},
         year = 2022,
        month = may,
       volume = {930},
       number = {1},
          eid = {80},
        pages = {80},
          doi = {10.3847/1538-4357/ac63ab},
archivePrefix = {arXiv},
       eprint = {2203.11953},
 primaryClass = {astro-ph.SR},
       adsurl = {https://ui.adsabs.harvard.edu/abs/2022ApJ...930...80M}
}

@ARTICLE{van_der_Marel_2013,
       author = {{van der Marel}, Nienke and {van Dishoeck}, Ewine F. and {Bruderer}, Simon and {Birnstiel}, Til and {Pinilla}, Paola and {Dullemond}, Cornelis P. and {van Kempen}, Tim A. and {Schmalzl}, Markus and {Brown}, Joanna M. and {Herczeg}, Gregory J. and {Mathews}, Geoffrey S. and {Geers}, Vincent},
        title = "{A Major Asymmetric Dust Trap in a Transition Disk}",
      journal = {Science},
         year = 2013,
        month = jun,
       volume = {340},
       number = {6137},
        pages = {1199-1202},
          doi = {10.1126/science.1236770},
archivePrefix = {arXiv},
       eprint = {1306.1768},
 primaryClass = {astro-ph.EP},
       adsurl = {https://ui.adsabs.harvard.edu/abs/2013Sci...340.1199V}
}

@ARTICLE{Meheut_2012,
       author = {{Meheut}, H. and {Meliani}, Z. and {Varniere}, P. and {Benz}, W.},
        title = "{Dust-trapping Rossby vortices in protoplanetary disks}",
      journal = {\aap},
         year = 2012,
        month = sep,
       volume = {545},
          eid = {A134},
        pages = {A134},
          doi = {10.1051/0004-6361/201219794},
archivePrefix = {arXiv},
       eprint = {1208.4947},
 primaryClass = {astro-ph.EP},
       adsurl = {https://ui.adsabs.harvard.edu/abs/2012A&A...545A.134M}
}

@ARTICLE{Zhu_2014a,
       author = {{Zhu}, Zhaohuan and {Stone}, James M. and {Rafikov}, Roman R. and {Bai}, Xue-ning},
        title = "{Particle Concentration at Planet-induced Gap Edges and Vortices. I. Inviscid Three-dimensional Hydro Disks}",
      journal = {\apj},
         year = 2014,
        month = apr,
       volume = {785},
       number = {2},
          eid = {122},
        pages = {122},
          doi = {10.1088/0004-637X/785/2/122},
archivePrefix = {arXiv},
       eprint = {1308.0648},
 primaryClass = {astro-ph.EP}


}

@article{Paardekooper_2010,
   title={VORTEX MIGRATION IN PROTOPLANETARY DISKS},
   volume={725},
   ISSN={1538-4357},
   url={http://dx.doi.org/10.1088/0004-637X/725/1/146},
   DOI={10.1088/0004-637x/725/1/146},
   number={1},
   journal={The Astrophysical Journal},
   publisher={American Astronomical Society},
   author={Paardekooper, Sijme-Jan and Lesur, Geoffroy and Papaloizou, John C. B.},
   year={2010},
   month=nov, pages={146--158} }

@ARTICLE{Ma_2025,
       author = {{Ma}, Xiaoyi and {Huang}, Pinghui and {Yu}, Cong and {Dong}, Ruobing},
        title = "{Vortex-induced Rings and Gaps within Protoplanetary Disks}",
      journal = {\apj},
         year = 2025,
        month = feb,
       volume = {979},
       number = {2},
          eid = {244},
        pages = {244},
          doi = {10.3847/1538-4357/ad9f2c},
archivePrefix = {arXiv},
       eprint = {2412.11507},
 primaryClass = {astro-ph.EP},
       adsurl = {https://ui.adsabs.harvard.edu/abs/2025ApJ...979..244M}
}

@article{Cazzoletti_2018,
   title={Evidence for a massive dust-trapping vortex connected to spirals: Multi-wavelength analysis of the HD 135344B protoplanetary disk},
   volume={619},
   ISSN={1432-0746},
   url={http://dx.doi.org/10.1051/0004-6361/201834006},
   DOI={10.1051/0004-6361/201834006},
   journal={Astronomy\&amp; Astrophysics},
   publisher={EDP Sciences},
   author={Cazzoletti, P. and van Dishoeck, E. F. and Pinilla, P. and Tazzari, M. and Facchini, S. and van der Marel, N. and Benisty, M. and Garufi, A. and Perez, L. M.},
   year={2018},
   month=nov, pages={A161} }

@ARTICLE{Wolfer_2025,
       author = {{W{\"o}lfer}, Lisa and {Barraza-Alfaro}, Marcelo and {Teague}, Richard and {Curone}, Pietro and {Benisty}, Myriam and {Fukagawa}, Misato and {Bae}, Jaehan and {Cataldi}, Gianni and {Czekala}, Ian and {Facchini}, Stefano and {Fasano}, Daniele and {Flock}, Mario and {Galloway-Sprietsma}, Maria and {Garg}, Himanshi and {Hall}, Cassandra and {Huang}, Jane and {Ilee}, John D. and {Izquierdo}, Andr{\'e}s F. and {Kanagawa}, Kazuhiro and {Lesur}, Geoffroy and {Longarini}, Cristiano and {Loomis}, Ryan A. and {Menard}, Francois and {Nath}, Anika and {Orihara}, Ryuta and {Pinte}, Christophe and {Price}, Daniel J. and {Rosotti}, Giovanni and {Stadler}, Jochen and {Wafflard-Fernandez}, Gaylor and {Winter}, Andrew J. and {Yen}, Hsi-Wei and {Yoshida}, Tomohiro C. and {Zawadzki}, Brianna},
        title = "{exoALMA. XVII. Characterizing the Gas Dynamics around Dust Asymmetries}",
      journal = {\apjl},
         year = 2025,
        month = may,
       volume = {984},
       number = {1},
          eid = {L22},
        pages = {L22},
          doi = {10.3847/2041-8213/adc42c},
archivePrefix = {arXiv},
       eprint = {2504.20023},
 primaryClass = {astro-ph.EP},
       adsurl = {https://ui.adsabs.harvard.edu/abs/2025ApJ...984L..22W}
}

@ARTICLE{Stammler_2019,
       author = {{Stammler}, Sebastian M. and {Dr{\.z}kowska}, Joanna and {Birnstiel}, Til and {Klahr}, Hubert and {Dullemond}, Cornelis P. and {Andrews}, Sean M.},
        title = "{The DSHARP Rings: Evidence of Ongoing Planetesimal Formation?}",
      journal = {\apjl},
         year = 2019,
        month = oct,
       volume = {884},
       number = {1},
          eid = {L5},
        pages = {L5},
          doi = {10.3847/2041-8213/ab4423},
archivePrefix = {arXiv},
       eprint = {1909.04674},
 primaryClass = {astro-ph.EP},
       adsurl = {https://ui.adsabs.harvard.edu/abs/2019ApJ...884L...5S}
}

@article{Li_2020,
   title={Planet-induced Vortices with Dust Coagulation in Protoplanetary Disks},
   volume={892},
   ISSN={2041-8213},
   url={http://dx.doi.org/10.3847/2041-8213/ab7fb2},
   DOI={10.3847/2041-8213/ab7fb2},
   number={2},
   journal={The Astrophysical Journal Letters},
   publisher={American Astronomical Society},
   author={Li, Ya-Ping and Li, Hui and Li, Shengtai and Birnstiel, Tilman and Dra̧żkowska, Joanna and Stammler, Sebastian},
   year={2020},
   month=mar, pages={L19} }

@ARTICLE{Fu_2014,
       author = {{Fu}, Wen and {Li}, Hui and {Lubow}, Stephen and {Li}, Shengtai and {Liang}, Edison},
        title = "{Effects of Dust Feedback on Vortices in Protoplanetary Disks}",
      journal = {\apjl},
         year = 2014,
        month = nov,
       volume = {795},
       number = {2},
          eid = {L39},
        pages = {L39},
          doi = {10.1088/2041-8205/795/2/L39},
archivePrefix = {arXiv},
       eprint = {1410.4196},
 primaryClass = {astro-ph.EP},
       adsurl = {https://ui.adsabs.harvard.edu/abs/2014ApJ...795L..39F}
}

@ARTICLE{Law_2023,
       author = {{Law}, Charles J. and {Teague}, Richard and {{\"O}berg}, Karin I. and {Rich}, Evan A. and {Andrews}, Sean M. and {Bae}, Jaehan and {Benisty}, Myriam and {Facchini}, Stefano and {Flaherty}, Kevin and {Isella}, Andrea and {Jin}, Sheng and {Hashimoto}, Jun and {Huang}, Jane and {Loomis}, Ryan A. and {Long}, Feng and {Romero-Mirza}, Carlos E. and {Paneque-Carre{\~n}o}, Teresa and {P{\'e}rez}, Laura M. and {Qi}, Chunhua and {Schwarz}, Kamber R. and {Stadler}, Jochen and {Tsukagoshi}, Takashi and {Wilner}, David J. and {van der Plas}, Gerrit},
        title = "{Mapping Protoplanetary Disk Vertical Structure with CO Isotopologue Line Emission}",
      journal = {\apj},
         year = 2023,
        month = may,
       volume = {948},
       number = {1},
          eid = {60},
        pages = {60},
          doi = {10.3847/1538-4357/acb3c4},
archivePrefix = {arXiv},
       eprint = {2212.08667},
 primaryClass = {astro-ph.EP},
       adsurl = {https://ui.adsabs.harvard.edu/abs/2023ApJ...948...60L}
}

@ARTICLE{Cimerman_2023,
       author = {{Cimerman}, Nicolas P. and {Rafikov}, Roman R.},
        title = "{Emergence of vortices at the edges of planet-driven gaps in protoplanetary discs}",
      journal = {\mnras},
         year = 2023,
        month = feb,
       volume = {519},
       number = {1},
        pages = {208-227},
          doi = {10.1093/mnras/stac3507},
archivePrefix = {arXiv},
       eprint = {2212.03062},
 primaryClass = {astro-ph.EP},
       adsurl = {https://ui.adsabs.harvard.edu/abs/2023MNRAS.519..208C}
}

@ARTICLE{Lovelace_1978,
       author = {{Lovelace}, R.~V.~E. and {Hohlfeld}, R.~G.},
        title = "{Negative mass instability of flat galaxies.}",
      journal = {\apj},
         year = 1978,
        month = apr,
       volume = {221},
        pages = {51-61},
          doi = {10.1086/156004},
       adsurl = {https://ui.adsabs.harvard.edu/abs/1978ApJ...221...51L}
}

@ARTICLE{Li_2000,
       author = {{Li}, H. and {Finn}, J.~M. and {Lovelace}, R.~V.~E. and {Colgate}, S.~A.},
        title = "{Rossby Wave Instability of Thin Accretion Disks. II. Detailed Linear Theory}",
      journal = {\apj},
         year = 2000,
        month = apr,
       volume = {533},
       number = {2},
        pages = {1023-1034},
          doi = {10.1086/308693},
archivePrefix = {arXiv},
       eprint = {astro-ph/9907279},
 primaryClass = {astro-ph},
       adsurl = {https://ui.adsabs.harvard.edu/abs/2000ApJ...533.1023L}
}

@ARTICLE{Lovelace_1999,
       author = {{Lovelace}, R.~V.~E. and {Li}, H. and {Colgate}, S.~A. and {Nelson}, A.~F.},
        title = "{Rossby Wave Instability of Keplerian Accretion Disks}",
      journal = {\apj},
         year = 1999,
        month = mar,
       volume = {513},
       number = {2},
        pages = {805-810},
          doi = {10.1086/306900},
archivePrefix = {arXiv},
       eprint = {astro-ph/9809321},
 primaryClass = {astro-ph},
       adsurl = {https://ui.adsabs.harvard.edu/abs/1999ApJ...513..805L}
}

@article{Lyra_2008,
   title={Standing on the shoulders of giants: Trojan Earths and vortex trapping
in low mass  self-gravitating protoplanetary disks of gas and solids},
   volume={493},
   ISSN={1432-0746},
   url={http://dx.doi.org/10.1051/0004-6361:200810797},
   DOI={10.1051/0004-6361:200810797},
   number={3},
   journal={Astronomy\&amp; Astrophysics},
   publisher={EDP Sciences},
   author={Lyra, W. and Johansen, A. and Klahr, H. and Piskunov, N.},
   year={2008},
   month=nov, pages={1125--1139}
}

@ARTICLE{Lin_2012,
       author = {{Lin}, Min-Kai},
        title = "{Vortex and spiral instabilities at gap edges in three-dimensional self-gravitating disc-satellite simulations}",
      journal = {\mnras},
         year = 2012,
        month = nov,
       volume = {426},
       number = {4},
        pages = {3211-3224},
          doi = {10.1111/j.1365-2966.2012.21955.x},
archivePrefix = {arXiv},
       eprint = {1205.4034},
 primaryClass = {astro-ph.EP},
       adsurl = {https://ui.adsabs.harvard.edu/abs/2012MNRAS.426.3211L}
}

@misc{Doi_2024,
      title={Asymmetric dust accumulation of the PDS 70 disk revealed by ALMA Band 3 observations}, 
      author={Kiyoaki Doi and Akimasa Kataoka and Hauyu Baobab Liu and Tomohiro C. Yoshida and Myriam Benisty and Ruobing Dong and Yoshihide Yamato and Jun Hashimoto},
      year={2024},
      eprint={2408.09216},
      archivePrefix={arXiv},
      primaryClass={astro-ph.EP},
      url={https://arxiv.org/abs/2408.09216}, 
}

@inbook{Testi_2014,
   title={Dust Evolution in Protoplanetary Disks},
   url={http://dx.doi.org/10.2458/azu_uapress_9780816531240-ch015},
   DOI={10.2458/azu_uapress_9780816531240-ch015},
   booktitle={Protostars and Planets VI},
   publisher={University of Arizona Press},
   author={Testi, L. and Birnstiel, T. and Ricci, L. and Andrews, S. and Blum, J. and Carpenter, J. and Dominik, C. and Isella, A. and Natta, A. and Williams, J. P. and Wilner, D. J.},
   year={2014} }

@article{Draine_2006,
   title={On the Submillimeter Opacity of Protoplanetary Disks},
   volume={636},
   ISSN={1538-4357},
   url={http://dx.doi.org/10.1086/498130},
   DOI={10.1086/498130},
   number={2},
   journal={The Astrophysical Journal},
   publisher={American Astronomical Society},
   author={Draine, B. T.},
   year={2006},
   month=jan, pages={1114--1120} }

@ARTICLE{Ragusa_2017,
       author = {{Ragusa}, Enrico and {Dipierro}, Giovanni and {Lodato}, Giuseppe and {Laibe}, Guillaume and {Price}, Daniel J.},
        title = "{On the origin of horseshoes in transitional discs}",
      journal = {\mnras},
         year = 2017,
        month = jan,
       volume = {464},
       number = {2},
        pages = {1449-1455},
          doi = {10.1093/mnras/stw2456},
archivePrefix = {arXiv},
       eprint = {1609.08159},
 primaryClass = {astro-ph.EP},
       adsurl = {https://ui.adsabs.harvard.edu/abs/2017MNRAS.464.1449R}
}

@ARTICLE{Mittal_2015,
       author = {{Mittal}, Tushar and {Chiang}, Eugene},
        title = "{Fast Modes and Dusty Horseshoes in Transitional Disks}",
      journal = {\apjl},
         year = 2015,
        month = jan,
       volume = {798},
       number = {1},
          eid = {L25},
        pages = {L25},
          doi = {10.1088/2041-8205/798/1/L25},
archivePrefix = {arXiv},
       eprint = {1412.2135},
 primaryClass = {astro-ph.EP},
       adsurl = {https://ui.adsabs.harvard.edu/abs/2015ApJ...798L..25M}
}

@ARTICLE{Marino_2015b,
       author = {{Marino}, S. and {Casassus}, S. and {Perez}, S. and {Lyra}, W. and {Roman}, P.~E. and {Avenhaus}, H. and {Wright}, C.~M. and {Maddison}, S.~T.},
        title = "{Compact Dust Concentration in the MWC 758 Protoplanetary Disk}",
      journal = {\apj},
         year = 2015,
        month = nov,
       volume = {813},
       number = {1},
          eid = {76},
        pages = {76},
          doi = {10.1088/0004-637X/813/1/76},
archivePrefix = {arXiv},
       eprint = {1505.06732},
 primaryClass = {astro-ph.EP},
       adsurl = {https://ui.adsabs.harvard.edu/abs/2015ApJ...813...76M}
}

@ARTICLE{Casassus_2019,
       author = {{Casassus}, Simon and {Marino}, Sebasti{\'a}n and {Lyra}, Wladimir and {Baruteau}, Cl{\'e}ment and {Vidal}, Mat{\'\i}as and {Wootten}, Alwyn and {P{\'e}rez}, Sebasti{\'a}n and {Alarcon}, Felipe and {Barraza}, Marcelo and {C{\'a}rcamo}, Miguel and {Dong}, Ruobing and {Sierra}, Anibal and {Zhu}, Zhaohuan and {Ricci}, Luca and {Christiaens}, Valentin and {Cieza}, Lucas},
        title = "{Cm-wavelength observations of MWC 758: resolved dust trapping in a vortex}",
      journal = {\mnras},
         year = 2019,
        month = mar,
       volume = {483},
       number = {3},
        pages = {3278-3287},
          doi = {10.1093/mnras/sty3269},
archivePrefix = {arXiv},
       eprint = {1805.03023},
 primaryClass = {astro-ph.SR},
       adsurl = {https://ui.adsabs.harvard.edu/abs/2019MNRAS.483.3278C}
}

@ARTICLE{Yen_Gu_2020,
       author = {{Yen}, Hsi-Wei and {Gu}, Pin-Gao},
        title = "{Kinematical Signs of Dust Trapping and Feedback in a Local Pressure Bump in the Protoplanetary Disk around HD 142527 Revealed with ALMA}",
      journal = {\apj},
         year = 2020,
        month = dec,
       volume = {905},
       number = {2},
          eid = {89},
        pages = {89},
          doi = {10.3847/1538-4357/abc55a},
archivePrefix = {arXiv},
       eprint = {2010.13990},
 primaryClass = {astro-ph.EP},
       adsurl = {https://ui.adsabs.harvard.edu/abs/2020ApJ...905...89Y}
}

@ARTICLE{Dullemond_2018,
       author = {{Dullemond}, Cornelis P. and {Birnstiel}, Tilman and {Huang}, Jane and {Kurtovic}, Nicol{\'a}s T. and {Andrews}, Sean M. and {Guzm{\'a}n}, Viviana V. and {P{\'e}rez}, Laura M. and {Isella}, Andrea and {Zhu}, Zhaohuan and {Benisty}, Myriam and {Wilner}, David J. and {Bai}, Xue-Ning and {Carpenter}, John M. and {Zhang}, Shangjia and {Ricci}, Luca},
        title = "{The Disk Substructures at High Angular Resolution Project (DSHARP). VI. Dust Trapping in Thin-ringed Protoplanetary Disks}",
      journal = {\apjl},
         year = 2018,
        month = dec,
       volume = {869},
       number = {2},
          eid = {L46},
        pages = {L46},
          doi = {10.3847/2041-8213/aaf742},
archivePrefix = {arXiv},
       eprint = {1812.04044},
 primaryClass = {astro-ph.EP},
       adsurl = {https://ui.adsabs.harvard.edu/abs/2018ApJ...869L..46D}
}

@article{Lyra_2013,
   title={STEADY STATE DUST DISTRIBUTIONS IN DISK VORTICES: OBSERVATIONAL PREDICTIONS AND APPLICATIONS TO TRANSITIONAL DISKS},
   volume={775},
   ISSN={1538-4357},
   url={http://dx.doi.org/10.1088/0004-637X/775/1/17},
   DOI={10.1088/0004-637x/775/1/17},
   number={1},
   journal={The Astrophysical Journal},
   publisher={American Astronomical Society},
   author={Lyra, Wladimir and Lin, Min-Kai},
   year={2013},
   month=aug, pages={17} }

@article{Casassus_2015,
   title={A COMPACT CONCENTRATION OF LARGE GRAINS IN THE HD 142527 PROTOPLANETARY DUST TRAP},
   volume={812},
   ISSN={1538-4357},
   url={http://dx.doi.org/10.1088/0004-637X/812/2/126},
   DOI={10.1088/0004-637x/812/2/126},
   number={2},
   journal={The Astrophysical Journal},
   publisher={American Astronomical Society},
   author={Casassus, Simon and Wright, Chris M. and Marino, Sebastian and Maddison, Sarah T. and Wootten, Al and Roman, Pablo and Perez, Sebastian and Pinilla, Paola and Wyatt, Mark and Moral, Victor and Menard, Francois and Christiaens, Valentin and Cieza, Lucas and Plas, Gerrit van der},
   year={2015},
   month=oct, pages={126} }

@ARTICLE{van_der_Marel_2015b,
       author = {{van der Marel}, N. and {Pinilla}, P. and {Tobin}, J. and {van Kempen}, T. and {Andrews}, S. and {Ricci}, L. and {Birnstiel}, T.},
        title = "{A Concentration of Centimeter-sized Grains in the Ophiuchus IRS 48 Dust Trap}",
      journal = {\apjl},
         year = 2015,
        month = sep,
       volume = {810},
       number = {1},
          eid = {L7},
        pages = {L7},
          doi = {10.1088/2041-8205/810/1/L7},
archivePrefix = {arXiv},
       eprint = {1508.01003},
 primaryClass = {astro-ph.EP},
       adsurl = {https://ui.adsabs.harvard.edu/abs/2015ApJ...810L...7V}
}

@misc{stadler_2026,
      title={The Circumbinary Disk of HD 34700A: I. CO gas kinematics indicate spirals, infall, and vortex motions}, 
      author={J. Stadler and M. Benisty and F. Zagaria and A. F. Izquierdo and J. Speedie and A. J. Winter and L. Wölfer and J. Bae and S. Facchini and D. Fasano and N. Kurtovic and R. Teague},
      year={2026},
      eprint={2601.15262},
      archivePrefix={arXiv},
      primaryClass={astro-ph.EP},
      url={https://arxiv.org/abs/2601.15262}, 
}

@ARTICLE{Boccaletti_2020,
       author = {{Boccaletti}, A. and {Di Folco}, E. and {Pantin}, E. and {Dutrey}, A. and {Guilloteau}, S. and {Tang}, Y.~W. and {Pi{\'e}tu}, V. and {Habart}, E. and {Milli}, J. and {Beck}, T.~L. and {Maire}, A.-L.},
        title = "{Possible evidence of ongoing planet formation in AB Aurigae. A showcase of the SPHERE/ALMA synergy}",
      journal = {\aap},
         year = 2020,
        month = may,
       volume = {637},
          eid = {L5},
        pages = {L5},
          doi = {10.1051/0004-6361/202038008},
archivePrefix = {arXiv},
       eprint = {2005.09064},
 primaryClass = {astro-ph.EP},
       adsurl = {https://ui.adsabs.harvard.edu/abs/2020A&A...637L...5B}
}

@ARTICLE{Tang_2017,
       author = {{Tang}, Ya-Wen and {Guilloteau}, Stephane and {Dutrey}, Anne and {Muto}, Takayuki and {Shen}, Bo-Ting and {Gu}, Pin-Gao and {Inutsuka}, Shu-ichiro and {Momose}, Munetake and {Pietu}, Vincent and {Fukagawa}, Misato and {Chapillon}, Edwige and {Ho}, Paul T.~P. and {di Folco}, Emmanuel and {Corder}, Stuartt and {Ohashi}, Nagayoshi and {Hashimoto}, Jun},
        title = "{Planet Formation in AB Aurigae: Imaging of the Inner Gaseous Spirals Observed inside the Dust Cavity}",
      journal = {\apj},
         year = 2017,
        month = may,
       volume = {840},
       number = {1},
          eid = {32},
        pages = {32},
          doi = {10.3847/1538-4357/aa6af7},
archivePrefix = {arXiv},
       eprint = {1704.02699},
 primaryClass = {astro-ph.GA},
       adsurl = {https://ui.adsabs.harvard.edu/abs/2017ApJ...840...32T}
}

@ARTICLE{Ragusa_2020,
       author = {{Ragusa}, Enrico and {Alexander}, Richard and {Calcino}, Josh and {Hirsh}, Kieran and {Price}, Daniel J.},
        title = "{The evolution of large cavities and disc eccentricity in circumbinary discs}",
      journal = {\mnras},
         year = 2020,
        month = dec,
       volume = {499},
       number = {3},
        pages = {3362-3380},
          doi = {10.1093/mnras/staa2954},
archivePrefix = {arXiv},
       eprint = {2009.10738},
 primaryClass = {astro-ph.EP},
       adsurl = {https://ui.adsabs.harvard.edu/abs/2020MNRAS.499.3362R}
}

@ARTICLE{Yoshida_2025,
       author = {{Yoshida}, Tomohiro C. and {Nomura}, Hideko and {Tsukagoshi}, Takashi and {Doi}, Kiyoaki and {Furuya}, Kenji and {Kataoka}, Akimasa},
        title = "{Dust Scattering Albedo at Millimeter Wavelengths in the TW Hya Disk}",
      journal = {\apj},
         year = 2025,
        month = feb,
       volume = {980},
       number = {1},
          eid = {50},
        pages = {50},
          doi = {10.3847/1538-4357/ad9f31},
archivePrefix = {arXiv},
       eprint = {2412.10731},
 primaryClass = {astro-ph.EP},
       adsurl = {https://ui.adsabs.harvard.edu/abs/2025ApJ...980...50Y}
}

@article{Longarini_2025,
   title={exoALMA. XII. Weighing and Sizing exoALMA Disks with Rotation Curve Modelling},
   volume={984},
   ISSN={2041-8213},
   url={http://dx.doi.org/10.3847/2041-8213/adc431},
   DOI={10.3847/2041-8213/adc431},
   number={1},
   journal={The Astrophysical Journal Letters},
   publisher={American Astronomical Society},
   author={Longarini, Cristiano and Lodato, Giuseppe and Rosotti, Giovanni and Andrews, Sean and Winter, Andrew and Stadler, Jochen and Izquierdo, Andres and Galloway-Sprietsma, Maria and Facchini, Stefano and Curone, Pietro and Benisty, Myriam and Teague, Richard and Bae, Jaehan and Barraza-Alfaro, Marcelo and Cataldi, Gianni and Czekala, Ian and Cuello, Nicolás and Fasano, Daniele and Flock, Mario and Fukagawa, Misato and Garg, Himanshi and Hall, Cassandra and Hammond, Iain and Hardiman, Caitlyn and Hilder, Thomas and Huang, Jane and Ilee, John D. and Isella, Andrea and Kanagawa, Kazuhiro and Lesur, Geoffroy and Loomis, Ryan A. and Menard, Francois and Orihara, Ryuta and Pinte, Christophe and Price, Daniel and Testi, Leonardo and Fernandez, Gaylor Wafflard- and Wölfer, Lisa and Yen, Hsi-Wei and Yoshida, Tomohiro C. and Zawadzki, Brianna},
   year={2025},
   month=apr, pages={L17} }

@article{Galloway_Sprietsma_2025,
   title={exoALMA. V. Gaseous Emission Surfaces and Temperature Structures},
   volume={984},
   ISSN={2041-8213},
   url={http://dx.doi.org/10.3847/2041-8213/adc437},
   DOI={10.3847/2041-8213/adc437},
   number={1},
   journal={The Astrophysical Journal Letters},
   publisher={American Astronomical Society},
   author={Galloway-Sprietsma, Maria and Bae, Jaehan and Izquierdo, Andres F. and Stadler, Jochen and Longarini, Cristiano and Teague, Richard and Andrews, Sean M. and Winter, Andrew J. and Benisty, Myriam and Facchini, Stefano and Rosotti, Giovanni and Zawadzki, Brianna and Pinte, Christophe and Fasano, Daniele and Barraza-Alfaro, Marcelo and Cataldi, Gianni and Cuello, Nicolás and Curone, Pietro and Czekala, Ian and Flock, Mario and Fukagawa, Misato and Gardner, Charles H. and Garg, Himanshi and Hall, Cassandra and Huang, Jane and Ilee, John D. and Kanagawa, Kazuhiro and Lesur, Geoffroy and Lodato, Giuseppe and Loomis, Ryan A. and Menard, Francois and Orihara, Ryuta and Price, Daniel J. and Wafflard-Fernandez, Gaylor and Wilner, David J. and Wölfer, Lisa and Yen, Hsi-Wei and Yoshida, Tomohiro C.},
   year={2025},
   month=apr, pages={L10} }

@article{Birnstiel_2024,
   title={Dust Growth and Evolution in Protoplanetary Disks},
   volume={62},
   ISSN={1545-4282},
   url={http://dx.doi.org/10.1146/annurev-astro-071221-052705},
   DOI={10.1146/annurev-astro-071221-052705},
   number={1},
   journal={Annual Review of Astronomy and Astrophysics},
   publisher={Annual Reviews},
   author={Birnstiel, Tilman},
   year={2024},
   month=sep, pages={157--202} }

@ARTICLE{Temmink_2025,
       author = {{Temmink}, Milou and {Booth}, Alice S. and {Leemker}, Margot and {van der Marel}, Nienke and {van Dishoeck}, Ewine F. and {Evans}, Lucy and {Keyte}, Luke and {Law}, Charles J. and {Notsu}, Shota and {{\"O}berg}, Karin and {Walsh}, Catherine},
        title = "{Characterising the molecular line emission in the asymmetric Oph-IRS 48 dust trap: Temperatures, timescales, and sub-thermal excitation}",
      journal = {\aap},
         year = 2025,
        month = jan,
       volume = {693},
          eid = {A101},
        pages = {A101},
          doi = {10.1051/0004-6361/202452175},
archivePrefix = {arXiv},
       eprint = {2411.12418},
 primaryClass = {astro-ph.EP},
       adsurl = {https://ui.adsabs.harvard.edu/abs/2025A&A...693A.101T}
}

@article{Quiroz_2022,
   title={Improving Planet Detection with Disk Modeling: Keck/NIRC2 Imaging of the HD 34282 Single-armed Protoplanetary Disk},
   volume={924},
   ISSN={2041-8213},
   url={http://dx.doi.org/10.3847/2041-8213/ac3e62},
   DOI={10.3847/2041-8213/ac3e62},
   number={1},
   journal={The Astrophysical Journal Letters},
   publisher={American Astronomical Society},
   author={Quiroz, Juan and Wallack, Nicole L. and Ren, Bin and Dong, Ruobing  and Xuan, Jerry W. and Mawet, Dimitri and Millar-Blanchaer, Maxwell A. and Ruane, Garreth},
   year={2022},
   month=jan, pages={L4} }

@article{Marino_2015a,
   title={SHADOWS CAST BY A WARP IN THE HD 142527 PROTOPLANETARY DISK},
   volume={798},
   ISSN={2041-8213},
   url={http://dx.doi.org/10.1088/2041-8205/798/2/L44},
   DOI={10.1088/2041-8205/798/2/l44},
   number={2},
   journal={The Astrophysical Journal},
   publisher={American Astronomical Society},
   author={Marino, S. and Perez, S. and Casassus, S.},
   year={2015},
   month=jan, pages={L44} }

@techreport{ALMA_THB_Cycle12_2025,
  title        = {ALMA Cycle 12 Technical Handbook},
  author       = {Cort{\'e}s, Paulo and Carpenter, John and Kameno, Seiji and Loomis, Ryan and Vila Vilar{\'o}, Baltasar and Immer, Katharina and Vlahakis, Catherine and Law, James and Stoehr, Felix and Saini, Kamaljeet and Hales, Antonio and Kneissl, Ruediger},
  year         = {2025},
  note         = {ALMA Doc. 12.3, ver. 1.0},
  doi          = {10.5281/zenodo.14933753},
  url          = {https://doi.org/10.5281/zenodo.14933753},
  institution  = {Atacama Large Millimeter/submillimeter Array (ALMA)}
}

@ARTICLE{Birnstiel_2018,
       author = {{Birnstiel}, Tilman and {Dullemond}, Cornelis P. and {Zhu}, Zhaohuan and {Andrews}, Sean M. and {Bai}, Xue-Ning and {Wilner}, David J. and {Carpenter}, John M. and {Huang}, Jane and {Isella}, Andrea and {Benisty}, Myriam and {P{\'e}rez}, Laura M. and {Zhang}, Shangjia},
        title = "{The Disk Substructures at High Angular Resolution Project (DSHARP). V. Interpreting ALMA Maps of Protoplanetary Disks in Terms of a Dust Model}",
      journal = {\apjl},
         year = 2018,
        month = dec,
       volume = {869},
       number = {2},
          eid = {L45},
        pages = {L45},
          doi = {10.3847/2041-8213/aaf743},
archivePrefix = {arXiv},
       eprint = {1812.04043},
 primaryClass = {astro-ph.SR},
       adsurl = {https://ui.adsabs.harvard.edu/abs/2018ApJ...869L..45B}
}

@ARTICLE{Zhu_2019,
       author = {{Zhu}, Zhaohuan and {Zhang}, Shangjia and {Jiang}, Yan-Fei and {Kataoka}, Akimasa and {Birnstiel}, Tilman and {Dullemond}, Cornelis P. and {Andrews}, Sean M. and {Huang}, Jane and {P{\'e}rez}, Laura M. and {Carpenter}, John M. and {Bai}, Xue-Ning and {Wilner}, David J. and {Ricci}, Luca},
        title = "{One Solution to the Mass Budget Problem for Planet Formation: Optically Thick Disks with Dust Scattering}",
      journal = {\apjl},
         year = 2019,
        month = jun,
       volume = {877},
       number = {2},
          eid = {L18},
        pages = {L18},
          doi = {10.3847/2041-8213/ab1f8c},
archivePrefix = {arXiv},
       eprint = {1904.02127},
 primaryClass = {astro-ph.EP},
       adsurl = {https://ui.adsabs.harvard.edu/abs/2019ApJ...877L..18Z}
}

@ARTICLE{Liu_2019,
       author = {{Liu}, Hauyu Baobab},
        title = "{The Anomalously Low (Sub)Millimeter Spectral Indices of Some Protoplanetary Disks May Be Explained By Dust Self-scattering}",
      journal = {\apjl},
         year = 2019,
        month = jun,
       volume = {877},
       number = {2},
          eid = {L22},
        pages = {L22},
          doi = {10.3847/2041-8213/ab1f8e},
archivePrefix = {arXiv},
       eprint = {1904.00333},
 primaryClass = {astro-ph.SR},
       adsurl = {https://ui.adsabs.harvard.edu/abs/2019ApJ...877L..22L}
}

@ARTICLE{Lovelace_2014,
       author = {{Lovelace}, R.~V.~E. and {Romanova}, M.~M.},
        title = "{Rossby wave instability in astrophysical discs}",
      journal = {Fluid Dynamics Research},
         year = 2014,
        month = aug,
       volume = {46},
       number = {4},
          eid = {041401},
        pages = {041401},
          doi = {10.1088/0169-5983/46/4/041401},
archivePrefix = {arXiv},
       eprint = {1312.4572},
 primaryClass = {astro-ph.SR},
       adsurl = {https://ui.adsabs.harvard.edu/abs/2014FlDyR..46d1401L}
}

@ARTICLE{Jennings_2020,
author = {{Jennings}, Jeff and {Booth}, Richard A. and {Tazzari}, Marco and {Rosotti}, Giovanni P. and {Clarke}, Cathie J.},
title = "{frankenstein: protoplanetary disc brightness profile reconstruction at sub-beam resolution with a rapid Gaussian process}",
journal = {\mnras},
year = 2020,
month = jul,
volume = {495},
number = {3},
pages = {3209-3232},
doi = {10.1093/mnras/staa1365},
archivePrefix = {arXiv},
eprint = {2005.07709},
primaryClass = {astro-ph.EP},
adsurl = {https://ui.adsabs.harvard.edu/abs/2020MNRAS.495.3209J}
}

@article{Vides_2025,
doi = {10.3847/1538-4357/ae0932},
url = {https://doi.org/10.3847/1538-4357/ae0932},
year = {2025},
month = {oct},
publisher = {The American Astronomical Society},
volume = {993},
number = {2},
pages = {178},
author = {Vides, Christina L. and Sallum, Steph and Eisner, Josh and Skemer, Andy and Murray-Clay, Ruth},
title = {NIRC2 Interferometric Imaging of the HD 34282 Transition Disk’s Small Grain Structure},
journal = {The Astrophysical Journal}
}

@ARTICLE{Dong_2017,
       author = {{Dong}, Ruobing and {Li}, Shengtai and {Chiang}, Eugene and {Li}, Hui},
        title = "{Multiple Disk Gaps and Rings Generated by a Single Super-Earth}",
      journal = {\apj},
         year = 2017,
        month = jul,
       volume = {843},
       number = {2},
          eid = {127},
        pages = {127},
          doi = {10.3847/1538-4357/aa72f2},
archivePrefix = {arXiv},
       eprint = {1705.04687},
 primaryClass = {astro-ph.EP},
       adsurl = {https://ui.adsabs.harvard.edu/abs/2017ApJ...843..127D}
}

@ARTICLE{Dong_2018_2,
       author = {{Dong}, Ruobing and {Li}, Shengtai and {Chiang}, Eugene and {Li}, Hui},
        title = "{Multiple Disk Gaps and Rings Generated by a Single Super-Earth. II. Spacings, Depths, and Number of Gaps, with Application to Real Systems}",
      journal = {\apj},
         year = 2018,
        month = oct,
       volume = {866},
       number = {2},
          eid = {110},
        pages = {110},
          doi = {10.3847/1538-4357/aadadd},
archivePrefix = {arXiv},
       eprint = {1808.06613},
 primaryClass = {astro-ph.EP},
       adsurl = {https://ui.adsabs.harvard.edu/abs/2018ApJ...866..110D}
}

@ARTICLE{Bae_2017,
       author = {{Bae}, Jaehan and {Zhu}, Zhaohuan and {Hartmann}, Lee},
        title = "{On the Formation of Multiple Concentric Rings and Gaps in Protoplanetary Disks}",
      journal = {\apj},
         year = 2017,
        month = dec,
       volume = {850},
       number = {2},
          eid = {201},
        pages = {201},
          doi = {10.3847/1538-4357/aa9705},
archivePrefix = {arXiv},
       eprint = {1706.03066},
 primaryClass = {astro-ph.EP},
       adsurl = {https://ui.adsabs.harvard.edu/abs/2017ApJ...850..201B}
}

@ARTICLE{ALMA_2015,
       author = {{ALMA Partnership} and {Brogan}, C.~L. and {P{\'e}rez}, L.~M. and {Hunter}, T.~R. and {Dent}, W.~R.~F. and {Hales}, A.~S. and {Hills}, R.~E. and {Corder}, S. and {Fomalont}, E.~B. and {Vlahakis}, C. and {Asaki}, Y. and {Barkats}, D. and {Hirota}, A. and {Hodge}, J.~A. and {Impellizzeri}, C.~M.~V. and {Kneissl}, R. and {Liuzzo}, E. and {Lucas}, R. and {Marcelino}, N. and {Matsushita}, S. and {Nakanishi}, K. and {Phillips}, N. and {Richards}, A.~M.~S. and {Toledo}, I. and {Aladro}, R. and {Broguiere}, D. and {Cortes}, J.~R. and {Cortes}, P.~C. and {Espada}, D. and {Galarza}, F. and {Garcia-Appadoo}, D. and {Guzman-Ramirez}, L. and {Humphreys}, E.~M. and {Jung}, T. and {Kameno}, S. and {Laing}, R.~A. and {Leon}, S. and {Marconi}, G. and {Mignano}, A. and {Nikolic}, B. and {Nyman}, L.-A. and {Radiszcz}, M. and {Remijan}, A. and {Rod{\'o}n}, J.~A. and {Sawada}, T. and {Takahashi}, S. and {Tilanus}, R.~P.~J. and {Vila Vilaro}, B. and {Watson}, L.~C. and {Wiklind}, T. and {Akiyama}, E. and {Chapillon}, E. and {de Gregorio-Monsalvo}, I. and {Di Francesco}, J. and {Gueth}, F. and {Kawamura}, A. and {Lee}, C.-F. and {Nguyen Luong}, Q. and {Mangum}, J. and {Pietu}, V. and {Sanhueza}, P. and {Saigo}, K. and {Takakuwa}, S. and {Ubach}, C. and {van Kempen}, T. and {Wootten}, A. and {Castro-Carrizo}, A. and {Francke}, H. and {Gallardo}, J. and {Garcia}, J. and {Gonzalez}, S. and {Hill}, T. and {Kaminski}, T. and {Kurono}, Y. and {Liu}, H.-Y. and {Lopez}, C. and {Morales}, F. and {Plarre}, K. and {Schieven}, G. and {Testi}, L. and {Videla}, L. and {Villard}, E. and {Andreani}, P. and {Hibbard}, J.~E. and {Tatematsu}, K.},
        title = "{The 2014 ALMA Long Baseline Campaign: First Results from High Angular Resolution Observations toward the HL Tau Region}",
      journal = {\apjl},
         year = 2015,
        month = jul,
       volume = {808},
       number = {1},
          eid = {L3},
        pages = {L3},
          doi = {10.1088/2041-8205/808/1/L3},
archivePrefix = {arXiv},
       eprint = {1503.02649},
 primaryClass = {astro-ph.SR},
       adsurl = {https://ui.adsabs.harvard.edu/abs/2015ApJ...808L...3A}
}

@ARTICLE{Andrews_2016,
       author = {{Andrews}, Sean M. and {Wilner}, David J. and {Zhu}, Zhaohuan and {Birnstiel}, Tilman and {Carpenter}, John M. and {P{\'e}rez}, Laura M. and {Bai}, Xue-Ning and {{\"O}berg}, Karin I. and {Hughes}, A. Meredith and {Isella}, Andrea and {Ricci}, Luca},
        title = "{Ringed Substructure and a Gap at 1 au in the Nearest Protoplanetary Disk}",
      journal = {\apjl},
         year = 2016,
        month = apr,
       volume = {820},
       number = {2},
          eid = {L40},
        pages = {L40},
          doi = {10.3847/2041-8205/820/2/L40},
archivePrefix = {arXiv},
       eprint = {1603.09352},
 primaryClass = {astro-ph.EP},
       adsurl = {https://ui.adsabs.harvard.edu/abs/2016ApJ...820L..40A}
}

@ARTICLE{Perez_2019,
       author = {{P{\'e}rez}, Sebasti{\'a}n and {Casassus}, Simon and {Baruteau}, Cl{\'e}ment and {Dong}, Ruobing and {Hales}, Antonio and {Cieza}, Lucas},
        title = "{Dust Unveils the Formation of a Mini-Neptune Planet in a Protoplanetary Ring}",
      journal = {\aj},
         year = 2019,
        month = jul,
       volume = {158},
       number = {1},
          eid = {15},
        pages = {15},
          doi = {10.3847/1538-3881/ab1f88},
archivePrefix = {arXiv},
       eprint = {1902.05143},
 primaryClass = {astro-ph.EP},
       adsurl = {https://ui.adsabs.harvard.edu/abs/2019AJ....158...15P}
}

@ARTICLE{Ono_2018,
       author = {{Ono}, Tomohiro and {Muto}, Takayuki and {Tomida}, Kengo and {Zhu}, Zhaohuan},
        title = "{Parametric Study of the Rossby Wave Instability in a Two-dimensional Barotropic Disk. II. Nonlinear Calculations}",
      journal = {\apj},
         year = 2018,
        month = sep,
       volume = {864},
       number = {1},
          eid = {70},
        pages = {70},
          doi = {10.3847/1538-4357/aad54d},
archivePrefix = {arXiv},
       eprint = {1807.08847},
 primaryClass = {astro-ph.EP},
       adsurl = {https://ui.adsabs.harvard.edu/abs/2018ApJ...864...70O}
}
\end{document}